\pdfoutput=1
\documentclass[conference,compsoc]{IEEEtran}

\usepackage[pdftex]{graphicx}
\usepackage{multirow}
\usepackage{xcolor}
\usepackage{etoolbox}
\usepackage{soul, color}
\usepackage{lipsum} 
\usepackage{balance} 
\usepackage{caption}
\usepackage{amssymb,mathtools}
\usepackage{nomencl}
\usepackage{booktabs}

\usepackage{subfigure}
\newcommand \ignore[1]{}

\newcommand \Ourdefense[1]{Our defense}
\newcommand \ourdefense[1]{our defense}
\newcommand \nameofdefense[0]{\texttt{FLShield}}

\newcommand{\blue}[1]{\textcolor{black}{#1}}
\newcommand{\bluesp}[1]{\textcolor{black}{#1}}

\usepackage{svg}
\usepackage[inline]{enumitem}

\ifCLASSOPTIONcompsoc
  \usepackage[nocompress]{cite}
\else
  \usepackage{cite}
\fi
\ifCLASSINFOpdf
  \usepackage[pdftex]{graphicx}
\else
\fi
\usepackage{amsmath}
\usepackage{algorithmic}

\begin{document}
%
\title{\Large \bf \nameofdefense{}: A Validation Based Federated Learning Framework to Defend Against Poisoning Attacks}



%
\author{\IEEEauthorblockN{Ehsanul Kabir,
Zeyu Song,
Md Rafi Ur Rashid and 
Shagufta Mehnaz
\IEEEauthorblockA{Penn State University\\ Email: \{ekabir, zzs5287, mur5028, smehnaz\}@psu.edu}
}}

\maketitle

\begin{abstract}
Federated learning (FL) is revolutionizing how we learn from data. With its growing popularity, it is now being used in many safety-critical domains such as autonomous vehicles and healthcare. Since thousands of participants can contribute in this collaborative setting, it is, however, challenging to ensure security and reliability of such systems. This highlights the need to design FL systems that are secure and robust against malicious participants’ actions while also ensuring high utility, privacy of local data, and efficiency. In this paper, we propose a novel FL framework dubbed as \nameofdefense{} that utilizes benign data from FL participants to validate the local models before taking them into account for generating the global model. This is in stark contrast with existing defenses relying on server’s access to clean datasets\textemdash an assumption often impractical in real-life scenarios and conflicting with the fundamentals of FL. We conduct extensive experiments to evaluate our \nameofdefense{} framework in different settings and demonstrate its effectiveness in thwarting various types of poisoning and backdoor attacks including a defense-aware one. \nameofdefense{} also preserves privacy of local data against gradient inversion attacks.
\end{abstract}

\section{Introduction}

Federated learning (FL) \cite{konevcny2016federated, fedavg} is a collaborative learning technique that allows multiple participants to jointly train a machine learning (ML) model  without sharing their own training data.
Each participant trains their local model and shares their local updates with the server which then aggregates the updates to generate a global model and sends it back to the participants.
This technique has two important features. Firstly, it can produce a model that is trained on a vast quantity of data provided by thousands of participants without compromising their data privacy. Secondly, decentralized training allows parallelization of the computation across multiple devices resulting in a significant speed boost in the training process.
Due to these attractive features, FL technique has been adopted to solve problems in many diverse domains such as autonomous vehicles\cite{fl_app_autonomous_vehicles, fl_app_autonomous_vehicles_blockchain},
industry 4.0~\cite{fl_app_autonomous_industry4_blockchain, fl_app_autonomous_industry4_privacy}, 
healthcare sectors~\cite{fl_app_digital_health, fl_app_predictive_health}, etc.

The growing adoption of FL systems necessitates a comprehensive investigation of their robustness and security. 
Security concerns arise from the involvement of numerous clients including potential adversaries aiming to disrupt the learning process. 
Designing an efficient and effective vetting scheme for such a large number of participants is challenging. 
Malicious clients may attack FL systems by sending local model updates trained on poisoned data~\cite{tolpegin2020data, Bag2018, bhagoji}
or by intentionally poisoning local models~\cite{modelpoisoning-gong, fl_attack_model_poisoning}.

Recently, two categories of attacks have garnered significant attention: \emph{poisoning} attacks that aim to corrupt the global model and \emph{inference} attacks targeting the theft of local participant data. 
Poisoning attacks are executed by malicious participants seeking to compromise the global model by sending malicious updates. 
Attackers may aim to produce a global model with poor performance on the primary task (untargeted poisoning attack)~\cite{modelpoisoning-gong, fl_attack_Inner_Product_Manipulation}
or one that performs poorly on a specific class (i.e., targeted poisoning attack)~\cite{tolpegin2020data}.
Another possible goal is to create a global model that behaves according to the attacker's intentions when presented with a designed trigger (e.g., backdoor attacks)~\cite{backdoor_2, Bag2018}.
A common defense strategy against poisoning attacks involves robust aggregation protocols capable of detecting and filtering malicious updates.
However, this detection and filtering strategy relies on the assumption that malicious updates can be distinguished from benign ones\textemdash an assumption that may not hold in cases where participant data distribution varies significantly.
As benign updates also diverge, separating malicious updates becomes more challenging. Moreover, a malicious participant might adapt their local training procedure to make their updates indistinguishable from benign ones. 
Therefore, defending against poisoning attacks necessitates a strategy that can validate local models to effectively detect and filter malicious updates.
Conversely, inference attacks can be initiated by inverting gradients to reconstruct client's training data from individual model updates~\cite{gradient_inversion_2020}.

\textbf{Challenges:} Although some existing works~\cite{fltrust} assume server's access to clean data for local model validation, such assumptions are often impractical\textemdash especially in privacy sensitive FL applications.
Hence, implementing a validation strategy introduces two challenges: how to obtain the validation data and how to perform validation without exposing the local models. 
The validation data should be representative of the participants' training data. In practice,  any dataset collected by the server may not sufficiently represent the diverse distribution of clients' data.
One naive solution is selecting a set of  clients as validators from the pool of all clients and using their data for validation of local models. However, sending local models to validators risks exposing them to gradient inversion attacks~\cite{gradient_inversion_2020, gradient_inversion_2021}.
Another alternate approach is to validate the global model instead and accepting or rejecting the global model based on the validation performance \cite{baffle}.
However, the averaged global model could be infected at each round and fail the validation test resulting in a denial-of-service.
Hence, neither local nor global models are ideal for validation. 
We name this challenge as \textit{validation subject dilemma}. 
Additionally, using unvetted FL participants as validators raises concerns as malicious validators can send false results, a challenge we refer to as the \textit{validation integrity dilemma}. 
These dilemmas present significant obstacles in employing validation as a defense strategy against poisoning attacks.

\textbf{Proposed defense:}
In this paper, we propose a novel validation-based defense framework against poisoning attacks,
\nameofdefense{},
that is able to simultaneously solve the validation subject and validation integrity dilemmas in different FL setups.
The proposed strategy solves the validation subject dilemma by using 
\textit{representative models}, which are crafted using multiple local models to ensure that the inference attack will be ineffective.
Each representative model is then validated, and based on the validation results, the local models that contributed to the best representative models are selected.
We propose two versions of \nameofdefense{} based on two representative model crafting algorithms: one with the use of 
\emph{clustering}
 and the other with a scheme that we refer to as
\emph{bijective} model generation.
The version that uses clustering-based algorithm is referred to as \nameofdefense{}$^{*}$ and the version that uses bijective representatives is referred to as \nameofdefense{}$^{\dagger}$.
Our framework also suggests which strategy to use based on the number of participants and the level of robustness required in the system.

We solve the validation integrity dilemma by using a validation protocol that uses a new  metric that we refer to as
loss impact per class (LIPC).
This metric can be applied to any classifier validation process in an FL system.
We rank the representative models based on the minimum value extracted from the LIPC of each model and show that the top $50\%$ ranked models are largely crafted using benign updates.
Additionally, we demonstrate that the LIPC scores evaluated by the benign validators on representative models are consistent even in a non-IID scenario.
We use a validation filtering mechanism in \nameofdefense{} to filter tailored  validation results by malicious validators and solve the validation integrity dilemma.

\textbf{Evaluating \nameofdefense{} against diverse attacks:}
We further show that the filtering technique of \nameofdefense{} enables it to defend against both untargeted and targeted poisoning attacks.
Different poisoning techniques, such as data poisoning or model poisoning, are equally ineffective because of the framework's focus on validating models before aggregating them.
Through experiments on datasets across multiple domains in IID and non-IID scenarios, we show that \nameofdefense{} can defend against poisoning attacks of three categories including a defense-aware one: untargeted poisoning, targeted label flipping, and backdoor attacks.

\textbf{Summary of contributions:}
In summary, this paper makes the following contributions:
\begin{itemize}[leftmargin=0pt, itemindent=15pt]
    \item We develop a novel FL poisoning defense framework, \nameofdefense{},  that employs an effective validation strategy to solve both the validation subject dilemma and  validation integrity dilemma. 
    \item We design a new metric named LIPC for validation purpose. We show that the LIPC metric is a reliable metric to perform validation.
    \item Through extensive experimentation, we show that our framework can defend against poisoning attacks of three categories: untargeted poisoning, targeted label flipping, and backdoor attacks without compromising the model performance. Moreover, our framework outperforms all existing defenses in terms of robustness and performance.
    \item We design \nameofdefense{}-aware attacks and show that \nameofdefense{} is robust against those.
    \item Finally, we show that the representative models shared for the purpose of validation can not be used to reconstruct training data of clients using state-of-the-art gradient inversion attacks.
\end{itemize}

\vspace{-0.1cm}
\section{Preliminaries}
\label{sec:preliminaries}
\vspace{-0.1cm}

\subsection{Federated Learning}
\vspace{-0.1cm}
Federated learning (FL) is a distributed paradigm enabling a set of clients $S$ to learn a shared global model, coordinated by a central server. 
Unlike conventional ML frameworks requiring centralized data, FL permits local model training without data sharing. 
In a training round $t \in [1,T]$, the server sends the global model $G_t$ to a randomly sampled  $S_t \subset S$ of size $n$. 
Clients $k \in S_t$ fine-tune $G_t$ using local data $D_k$ and send updates $\omega_{k}^t$ to the server which then aggregates them to update the global model $G_{t+1}$.
FL minimizes this objective function:
$\min_{\omega}\sum_{k \in S} p_{k}F_{k}(\omega)$
where $F_{k}$ represents the local objective for client $k$, $\omega$ represents the model parameters and $p_{k}$ denotes their relative impact/weight.

\subsection{FL Poisoning Attacks}
\subsubsection{Untargeted Poisoning Attack}
Malicious FL participants may submit tampered updates to indiscriminately poison the global model and reduce the main accuracy (MA) of the FL task. This attack aims to minimize:
\begin{equation}
    \label{eq:acc}
    \text{{MA}} = \underset{(x,y) \sim \mathcal{D}}{\mathbb{E}} \left[ Pr(G(x) = y) \right]
\end{equation}
where $\mathcal{D}$ is the data distribution for the learning task, and $y$ is the expected output for input $x$.

\subsubsection{Targeted Poisoning Attack}
\blue{
Targeted poisoning attacks aim to impair global model performance on specific samples, while maintaining high accuracy elsewhere, making detection challenging.
For these attacks, the adversary aims to maximize misclassification rate (MR) on the targeted subset of data:
$$ \text{MR} = \underset{\substack{(x,y) \sim \mathcal{D}_{t}\\ \mathcal{D}_{t} \subset \mathcal{D}}}{\mathbb{E}} \left[ Pr(G(x) \neq y) \right]$$
where $\mathcal{D}_{t}$ are the samples targeted by the adversary.
}

\vspace{0.1cm}
\noindent\textbf{Targeted Label Flipping Attack}.
\blue{
This is a subclass of targeted poisoning attacks where the adversary manipulates the global model to misclassify samples from a specific class (i.e., source class) to a predetermined poison class (i.e., target class) \cite{tolpegin2020data}.
For these attacks, the adversary aims at two metrics: minimizing global model's recall (RCL) on source class and maximizing the attack success rate (ASR) which measures successful misclassification to the target class:
\begin{equation}
    \label{eq:recall_asr}
    \begin{aligned}
\text{RCL} &= \underset{\substack{(x,y) \sim \mathcal{D}\\ y = y_{s} }}{\mathbb{E}} \left[ Pr(G(x) = y) \right] \\
\text{ASR} &= \underset{\substack{(x,y) \sim \mathcal{D}\\ y = y_{s}}}{\mathbb{E}} \left[ Pr(G(x) = y_{t}) \right]
\end{aligned}
\end{equation}
where $y_{s}$ is the source class  and $y_{t}$ is the target class.
}

\vspace{0.1cm}
\noindent\textbf{Backdoor Attack}.
\blue{
In backdoor attacks, the global model learns a malicious sub-task alongside the original task, thereby misclassifying the inputs containing some attacker-chosen patterns. The attack aims to maximize the accuracy of this malicious sub-task also termed as backdoor accuracy (BA):
\begin{equation}
    \label{eq:asr_backdoor}
\text{BA} = \underset{\substack{(x,y) \sim \mathcal{D}\\ y \neq y'}}{\mathbb{E}} \left[ Pr(G(t(x)) = y') \right]
\end{equation}
Here, $t(.)$ is the backdoor trigger function and the pre-determined class $y'$ is the target class.
}

\subsection{FL Privacy Attacks}
While FL strives to protect privacy by keeping data at local devices, it still remains possible for attackers to launch privacy attacks if they gain access to the gradients \cite{gradient_inversion_2020, gradient_inversion_2021, wei2020framework}. These attacks are generally formulated as optimization problems and try to reconstruct $x_r$ that approximates the original instance $x$.

\section{Threat Model}
\label{sec:threat_model}

\noindent\textbf{Nature of the adversary:} 
We assume that the adversary compromises a subset of the clients through Sybil attacks by creating multiple fake clients to gain control over the FL system.
The Sybils have data sampled from the same distribution as the benign clients and they are coordinated by the adversary.
Each Sybil leaves as soon as it performs the attack, and then a new Sybil is created.
We assume that the FL coordinating server is honest.
\nameofdefense{} is capable of preserving privacy against an honest-but-curious server through the use of secure two-party computation which we present in Appendix \ref{sec:privacy}.

\noindent\textbf{Goal of the adversary:}
The adversary has the following objectives:
\begin{enumerate*}[label=(\arabic*)]
\item poison the global model (poisoning objective) and
\item reconstruct benign clients' training data (inference objective).
\end{enumerate*}
To achieve the poisoning objective, the adversary launches a targeted/untargeted poisoning attack, and the success is measured using metrics ASR/BA (targeted) or MA (untargeted).
The adversary aims to reconstruct the clients' training data by launching a gradient inversion attack on model updates.
Generally, the adversary has access to only global model updates. However, according to the design of \nameofdefense{}, when Sybils are selected as validators in a round, the adversary also gets access to the representative model updates.
Hence, in the threat model, we assume that the adversary tries to invert representative model updates.

\vspace{-0.1cm}
\section{Challenges and Key Insights of \nameofdefense{}}
\vspace{-0.1cm}
\subsection{Challenges}

In the ML domain, validation is the process of verifying a model's potential efficacy in the test phase by evaluating its performance on a validation dataset. 
In centralized learning, this set is typically a portion of the training dataset not utilized for training. 
Contrarily, in the context of FL, the participating clients are the data sources. Hence, a naive approach in FL could be using a portion of clients' data for validation. 
However, privacy concerns that led the invention of FL prevent the server from directly accessing clients' data. Although some existing works~\cite{fltrust} assume server's access to clean data, such assumptions are often impractical\textemdash especially in privacy sensitive FL applications.
Thus, the server must resort to the clients for validation of models. 
Now, resorting to clients could be insecure as malicious clients can submit falsified validation results to the server and disrupt the validation process. 
An effective defense mechanism must be able to discern genuine validation results from fraudulent ones. Moreover, malicious validator clients can attempt inference attacks on models received for validation\textemdash further complicating the challenges faced by \nameofdefense{}. Hence, \nameofdefense{}'s validation mechanism should only transmit models to validators that are not vulnerable to inference attacks. Addressing these two primary challenges is crucial for ensuring an effective defense.

\vspace{-0.1cm}
\subsubsection{Challenge 1: Validation Subject Dilemma}
This dilemma alludes to the challenge of selecting/computing appropriate models for validation in the FL setting, balancing the need for accuracy and privacy. 
The chosen validation model must be resilient to inference attacks while maintaining sufficient accuracy for validation purposes.

\vspace{-0.1cm}
\subsubsection{Challenge 2: Validation Integrity Dilemma}
This dilemma pertains to the challenge of ensuring the integrity of the validation results in the FL setting, particularly in discerning between genuine and falsified validation results. 
The necessity for verification of these validation results stems from the fact that the clients performing the validation could be controlled by the adversary.

\vspace{-0.1cm}
\subsection{Solving The Validation Subject Dilemma}
\label{sec:validation-subject-dilemma}
To address this dilemma, we introduce the concept of representative models which are fusions of multiple local model updates. 
We utilize two strategies to generate such representative models:
\begin{enumerate*}[label=(\arabic*)]
\item \emph{bijective representative models}: each local model update is treated as a base update and then accompanied by additional contributions from other local updates weighted according to their similarity to the base update.
\item \emph{cluster representative models}: model updates are clustered into multiple groups, and updates within each group are aggregated to form a representative model.
\end{enumerate*}

By aggregating local model updates trained on the datasets of multiple clients, representative models yield more generalized results and are thus more suitable for validation.
The integration of multiple local updates also thwarts privacy attacks launched by adversary-controlled malicious validators as it prevents successful reconstruction of training data. 
We perform gradient inversion attack on these representative models and observe that these attacks are unsuccessful in reconstructing any image resembling the clients' training data (details later in section \ref{sec:gradient_inversion_attack}).
Both of the generation strategies mentioned above demonstrate a high probability of merging benign updates with each other, and vice versa (section \ref{ablation:bijective}).

\subsubsection{Bijective Representative Models}
\label{subsubsec:bijective}
The fundamental concept of bijective representative models involves utilizing one local model update as the foundation and incorporating input from other updates based on their similarity to the foundation (i.e., the base model).
Formally, let $\mathcal{E}$ be the representative model, $\omega$ be the base update, and $\mathcal{U}$ be the set of all updates. The bijective representative model is defined as:

\begin{equation}
\label{eqn:bijective_ensembling}
    \mathcal{E} = G_t + (1-\tau) \omega + \tau \frac{\sum_{u \in \mathcal{U}-\{\omega\}}\left(s(\omega, u) \times u \times \frac{|\omega|}{|u|} \right)}{\sum_{u \in \mathcal{U}-\{\omega\}}(s(\omega, u) \times \frac{|\omega|}{|u|})}
\end{equation}
where $s(\omega, u)$ denotes the similarity between base update $\omega$ and another update $u$, and $\tau$ denotes the proportionate contribution ($0 < \tau < 1$) from other model updates to the representative model.
We further name the term associated with $\tau$ as the \emph{sibling contribution}.
Owing to the one-to-one relationship between each representative model and local model, we coin the term \emph{bijective} representative model.
In section \ref{subsubsec:Bijective-design}, we discuss the requirements for the similarity function $s(\omega, u)$ and our  approach to fulfill them.
Sibling contribution is incorporated into the representative model to enhance its generalizability.
This is because individual models trained on client-specific data subsets have weaker generalization than representative models which are composed of models trained on multiple clients’ datasets.
However, for \nameofdefense{} to mount a successful defense, two conditions must be fulfilled: (1) the sibling contribution need to be balanced to ensure the representative models achieve sufficient generalization without compromising the preeminent role of the base model, and (2) the benign representative model must contain a lesser proportion of contribution from malicious updates compared to those originating from other benign updates.
In sections \ref{ablation:bijective} and \ref{ablation:validation}, we demonstrate how fulfilling these conditions lead to a successful defense.

\subsubsection{Cluster Representative Models}
\label{subsubsec:cluster}
The technique for generating cluster representative models consists of two steps: (1) dynamically clustering clients based on their model updates, and (2) averaging the updates within each cluster to create a representative model.
Existing defenses claim that benign and malicious clients can be separated by clustering on the model updates~\cite{shen2016, Flame, li2021}.
However, in non-IID scenarios, distinguishing them becomes challenging as data distribution shifts can cause benign model updates to seem more divergent than malicious ones.
Despite this, we recognize the value of clustering as a useful tool.
The non-IID data prevalent among FL clients necessitates a clustering algorithm that dynamically adjusts the number of clusters. 
The dynamic clustering coupled with a quality evaluation metric is expected to group clients in a way that minimizes the intermingling of benign and malicious updates.
The veracity of the aforementioned claim is evidenced by experimental findings presented later in section \ref{ablation:miscellaneous}.
This fact holds significance as it suggests that we can distinguish benign from malicious representative models during validation, presuming benign representative models surpass malicious ones in performance.
Consequently, the combination of clustering and validation mechanisms not only aids in the separation of benign and malicious clients but also ensures that the selected representative models contribute positively to the global model\textemdash enhancing its overall performance and robustness against attacks.
The specifics of our clustering algorithm along with its associated hyperparameters are discussed in detail in section \ref{subsubsec:cluster-design}.

\noindent\textbf{Suitable Use-cases}.
FL systems host hundreds to tens of thousands of participants.
Bijective representative model generation, while computationally demanding, ensures individual evaluation, making it apt for smaller systems. 
Conversely, the clustering-based approach, offering higher performance throughput, may blend benign and malicious updates in smaller cohorts, thus proving more suitable for larger systems.

\vspace{-0.1cm}
\subsection{Solving the Validation Integrity Dilemma}
\label{subsec:validation_integrity}
In \nameofdefense{}, local clients act as validators for representative models, assess them and report validation results to the server. 
Nevertheless, verifying the integrity of these validation results is challenging. 
Variations in data distribution among clients could lead to different validation outcomes from different benign validators for the same representative model\textemdash making the identification of false results based on discrepancies from true results potentially challenging.
Therefore, we introduce a new validation metric, loss impact per class (LIPC), which is computed class-wise to reduce the influence of diverse data distributions on validation results. 
Our experiments reveal that employing LIPC in the validation process causes honest validation results to be more consistent with each other (section  \ref{ablation:validation}). We formally define the LIPC metric as follows:

\vspace{-0.2cm}
{\fontsize{8}{11}\selectfont
\begin{equation}
\label{eq:L}
\mathcal{L}(\mathcal{E}, v) = \left[ \underset{(x,y) \in D_{v},y=i}{Mean}L(G, (x,y)) -  L(\mathcal{E}, (x,y)) | \quad i \in [1, c] \right]
\end{equation}
\begin{equation}
\label{eq:ML}
\mathcal{M}(\mathcal{E}) = \left[ \mathcal{L}(\mathcal{E}, G, v_i)^T | \quad i \in [1, k] \right]^T
\end{equation}
\begin{equation}
\label{eq:NL}
\mathcal{N}(v) = \left[ \mathcal{L}(\mathcal{E}_i, G, v)^T | \quad i \in [1, m] \right]^T
\end{equation}
}
$\mathcal{L}$, a vector of class-wise loss value differences between the representative model and the prior round's global model, focuses on loss impact rather than raw loss values (due to the incremental nature of FL). 
$D_{v}$ represents the validation dataset of validator $v$, $G$ denotes the global model from  previous round, $\mathcal{E}$ is the representative model under validation, $L$ is the cross-entropy loss function, and $c$ signifies the number of classes.
$\mathcal{M}$ is a matrix for $\mathcal{E}$ containing all validators' $\mathcal{L}$ vectors.
$\mathcal{N}$ is a matrix from $v$ containing $\mathcal{L}$ vectors computed for all representative models.
$k$ and $m$ represent the number of validators and representative models, respectively.

For a validation sample $(x,y)$ and a representative model $\mathcal{E}$, the cross-entropy loss is given by $-\log(\mathcal{E}(x)[y])$, which represents the logit output for the correct class. 
A stable and benign neural network should yield similar logit values for all samples within a class, while a poisoned network is expected to produce varying logits for the source class. 
By computing the representative model's $\mathcal{L}$ for each class, we anticipate similar $\mathcal{N}$ from benign validators \bluesp{even in non-IID scenarios}.
As shown later in section \ref{ablation:validation} \bluesp{and \ref{subsec:non_iid}}, $\mathcal{N}$ from benign validators are indeed close to one another, supporting this hypothesis.

We perform thorough testing in various distribution scenarios to confirm the effectiveness of LIPC metric in ensuring validation integrity. 
We introduce label skew and feature skew in validators' validation data (section \ref{sec:overall_eval}) as well as vary false validation result injection strategies (section \ref{sec:mal_val_attack}). 
In all cases, use of LIPC enabled \nameofdefense{} to detect false validation results effectively.  
Moreover, using LIPC leads to the best performance amongst the metrics we consider later in section \ref{ablation:validation}.
These experiments substantiate the robustness and efficacy of our proposed validation approach.

\vspace{-0.1cm}
\section{Design and Implementation of \nameofdefense{}}

\begin{figure*}[ht]
    \centering
    \includegraphics[width=0.8\textwidth]
    {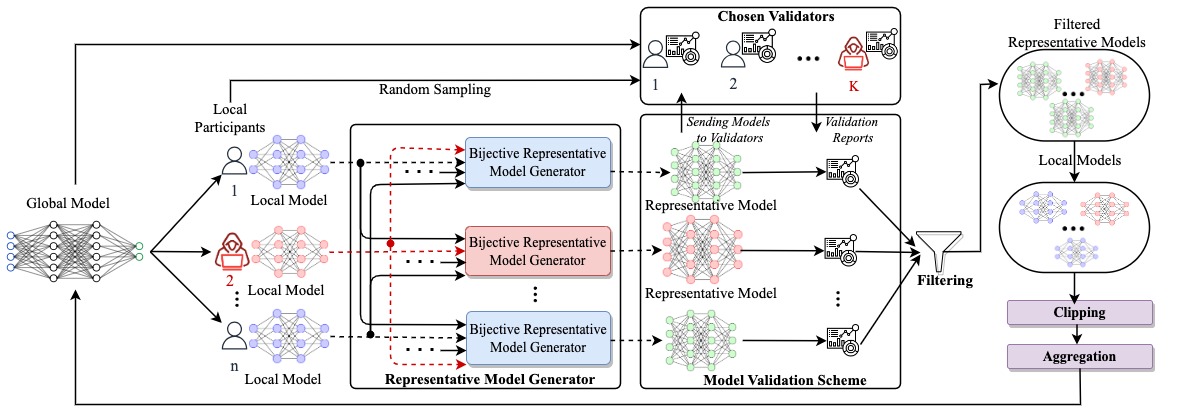}
    \caption{\nameofdefense{} framework with bijective representative models.}
    \label{fig:algorithm}
\end{figure*}

\begin{figure}[t]
 \scriptsize
    \begin{algorithmic}[1]
    \caption{\nameofdefense{} Algorithm}
    \label{alg:main}
    \REQUIRE{
        $G_0, T \quad \triangleright$ $G_0$ is the initial global model, $T$ is the number of training iterations
    }
    \ENSURE{
         $G_T \quad \triangleright$ $G_T$ is the final global model
    }
    \FOR{each training iteration t in [1, T]}{
        \STATE $\triangleright$ Local Training
        
        \FOR{each client i in [1, n]}{
            \STATE $\omega_i \gets ClientUpdate()$
        }
        \ENDFOR

        \STATE $M, (\mathcal{E}_1, \dots, \mathcal{E}_m) \gets RepresentativeGen(\omega_{1:n})$ 
        $\quad \triangleright M: [1, n] \times [1, m]$ is the mapping function between local model updates and representative models $\mathcal{E}_1, \dots, \mathcal{E}_m$, for bijective representative it is identity mapping
        \STATE $\mathcal{M}_1, \dots, \mathcal{M}_m \gets FedValidation(G_t, \mathcal{E}_1, \dots, \mathcal{E}_m) \quad \triangleright \mathcal{M}_1, \dots, \mathcal{M}_m$ are the $\mathcal{M}$ report for each representative model by the unfiltered validators
        \STATE $\mathbb{I}_{r} \gets Filtering(\mathcal{M}_1, \dots, \mathcal{M}_m)$
        $\quad \triangleright \mathbb{I}_{r}$ is the set of indices of the representative models that are accepted by the filtering scheme
        \STATE $\mathbb{I} \gets \{M_i : \forall i \in \mathbb{I}_{r} \}$
        $\quad \triangleright \mathbb{I}$ is the set of indices of the local models that are accepted by the validation scheme
        \STATE $\tilde{\omega}_{1:|\mathbb{I}|} \gets Clipping(\{ \omega_i : \forall i \in \mathbb{I} \})$
        $\quad \triangleright \tilde{\omega}_{1:|\mathbb{I}|}$ is the set of clipped local model updates
        \STATE $G_{t+1} \gets Aggregation(G_t, \tilde{\omega}_{1:|\mathbb{I}|})$
    }
    \ENDFOR
    \RETURN $G_T$
    \end{algorithmic}
\end{figure}

\nameofdefense{}'s  pipeline in each FL training round consists of the following steps:
\begin{enumerate*}[label=(\arabic*)]
\item Clients train local models and submit them to the server.
\item Server generates representative models from local model updates.
\item Validators evaluate the representative models and report $\mathcal{N}$ to the server.
\item Server filters outlier $\mathcal{N}$s.
\item Server selects top $50\%$ representative models after ranking them based on the projection of $\mathcal{M}$ to a scalar value.
\item Local models in accepted representative models are selected.
\item Selected local model updates undergo norm-bounded clipping.
\item Clipped local model updates are averaged to obtain the new global model.
\end{enumerate*}

\nameofdefense{} is designed to be flexible and adaptable to various FL systems, considering factors such as client numbers, client-server relationships, and computational resources. 
We propose two algorithms for representative model generation: \emph{bijective} and \emph{cluster}, detailed in sections~\ref{subsubsec:Bijective-design} and ~\ref{subsubsec:cluster-design}, respectively. 

Figure \ref{fig:algorithm} illustrates one instantiation of the \nameofdefense{} defense algorithm using  \emph{bijective} representative model generator. 
Other defense algorithm components remain abstracted, providing a high-level overview of  \nameofdefense{}. 
The general framework of \nameofdefense{} is outlined in the Algorithm presented in Figure \ref{alg:main}.
\nameofdefense{} is divided into five components:
\begin{enumerate*}
    \item representative model generator,
    \item model validation,
    \item filtering,
    \item clipping, and
    \item aggregation.
\end{enumerate*}
We discuss each of these components in the following sections.

\subsection{Representative Model Generator}
\label{subsec:representative_design}
We mentioned two strategies for representative model generation in section \ref{sec:validation-subject-dilemma}.
In this section, we explain the implementation details.

\vspace{-0.1cm}
\subsubsection{Bijective Representative Model Generator}
\label{subsubsec:Bijective-design}
We provide an implementation of the bijective representative generation method in the Algorithm presented in Figure \ref{alg:Bijective}. 
This approach constructs a representative model for each client using their local update as the base, and adding sibling contribution using equation \ref{eqn:bijective_ensembling}. 
The similarity function must satisfy two conditions:
\begin{enumerate*}[label=(\arabic*)]
    \item higher contribution from updates with similar direction
    \item zero contribution from updates with opposing direction.
\end{enumerate*}
We have chosen a combination of ReLU and cosine similarity as the similarity function.
ReLU's rectification meets the second criterion and cosine similarity's constrained output range meets the first criterion.

\subsubsection{Cluster Representative Model Generator}
\label{subsubsec:cluster-design}

This generator using the Algorithm presented in Figure \ref{alg:clustering} applies K-Means clustering to local model updates. 
To identify the optimal cluster count, the algorithm utilizes  silhouette score \cite{rousseeuw1987silhouettes}, a popular metric for optimal cluster determination, with min-max limits set by the system designer.
After determining the cluster count, it generates representative models by averaging the updates in each cluster. Finally, it returns a client-cluster mapping and the representative models. 

\vspace{-0.1cm}
\subsection{Model Validation}
The implementation of the model validation unit is detailed in the Algorithm presented in Figure \ref{alg:validation_new}. 
For each representative model, $k$ validators are randomly selected from the clients.
The outer loop (line 2) iterates over representative models, allowing parallelization if many clients are available, leading to lower computational cost and higher throughput. 
In systems with fewer clients, these clients typically possess more computational resources, allowing one validator to evaluate multiple representative models despite high-latency communication.

\subsubsection{$\mathcal{L}$ Calculation by Validator}
The Algorithm presented in Figure \ref{alg:lipc} illustrates the $\mathcal{L}$ calculation by the validator, with lower and upper bounds ($n_1$ and $n_2$) set for the number of validation samples for each class. 
Validators having fewer than $n_1$ samples are suggested to leave the value empty for that class. 
On the other hand, employing over $n_2$ samples might slow down the validation process without providing substantial defense benefits. Hence, validators are discouraged from using more than $n_2$ samples.
Although validators might deviate from these specifications, adherence is expected to maintain consistency and prevent result discarding.

\subsubsection{Filtering $\mathcal{N}$ anomalies by the Server}
The server utilizes IterativeImputation algorithm \cite{fancyimpute} (which outperforms others as shown in section \ref{ablation:miscellaneous}) to fill missing values in $\mathcal{L}$ scores before applying an outlier detection algorithm to $\mathcal{N}$ matrices to filter potential malicious validation reports. 
We use elliptic envelope for outlier detection. However, according to our observation, the choice of the algorithm does not impact performance at all as long as it is a robust one (section \ref{sec:outlier_detector_experiment}).
We further show that the absence of the outlier detector leads to failure under \nameofdefense{}-aware attacks (section \ref{ablation:miscellaneous}).

\subsection{Filtering}
\vspace{-0.1cm}
The filtering unit outlined in the Algorithm presented in Figure \ref{alg:filtering} is responsible for filtering malicious representative models based on their $\mathcal{M}$ matrix.
We perform the following steps on each $\mathcal{M}$:
\begin{enumerate*}[label=(\arabic*)]
    \item The matrix is averaged across the first dimension i.e. all the $\mathcal{L}$ vector reported by the unfiltered validators for the corresponding representative model are averaged
    \item The minimum value of the average $\mathcal{L}$ vector is extracted.
\end{enumerate*}
Afterwards, we examine the minimum value for each representative model, selecting the top $50\%$ based on these scores across all classes. 
This choice reflects our threat model's assumption that malicious clients constitute less than 50\% of the total clients. 
Untargeted poisoning attacks generally influence the $\mathcal{L}$ vectors across all classes, whereas targeted attacks predominantly impact one class. 
As a result, utilizing the minimum validation scores proves to be an effective measure for detecting both targeted and untargeted poisoning attacks.
Upon selecting the representative models, we utilize the mapping to identify the local model updates that contributed to these chosen representative models (line 9, \nameofdefense{} Algorithm).
In the remainder of the literature, we refer to the minimum of the $\mathcal{L}$ vector as the $\mathcal{L}$ score and the class-by-class evaluation as $\mathcal{L}$ vector.

\vspace{-0.1cm}
\subsection{Clipping}
\label{sec:clipping}
In FL, malicious clients can send disproportionately large model updates to unfairly and detrimentally influence the global model. 
To counter this issue, \nameofdefense{} incorporates a clipping technique similar to FLAME \cite{Flame}.

\subsection{Aggregation}
To obtain the new global model, we average the selected local model updates and combine them with  previous round's global model.

\begin{figure*}[t]
    \centering
    \subfigure{
        \includegraphics[width=0.3\textwidth]{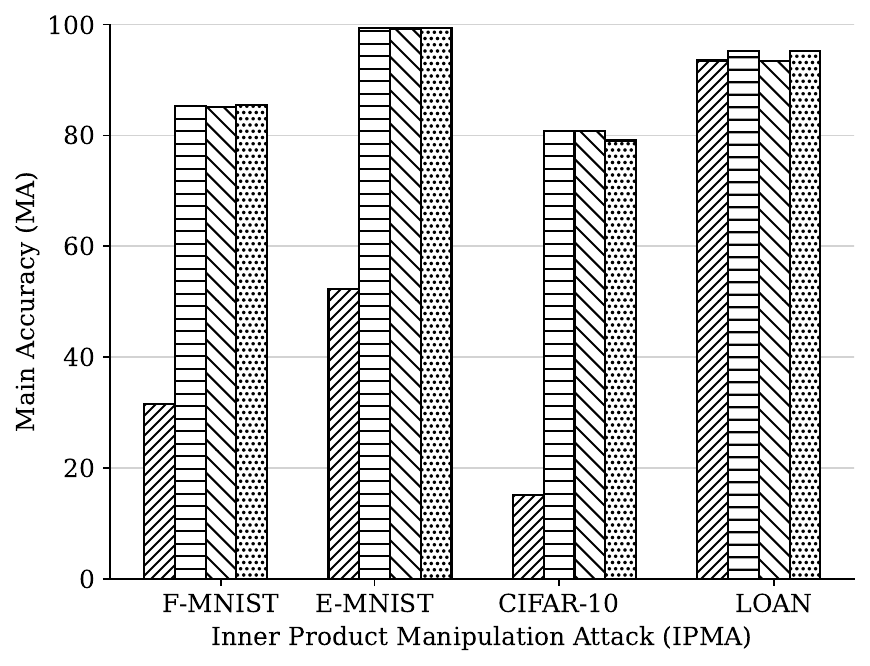}
        \label{fig:graph3_ipm}
    }
    \hspace{-8pt}
    \subfigure{
        \includegraphics[width=0.3\textwidth]{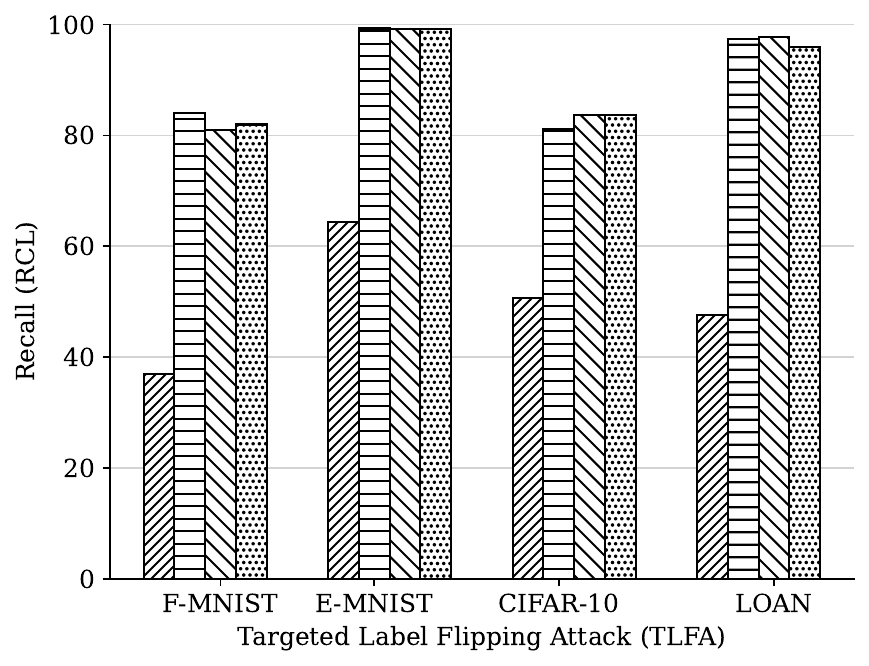}
        \label{fig:graph1_tlf}
    }
    \subfigure{
        \includegraphics[width=0.3\textwidth]{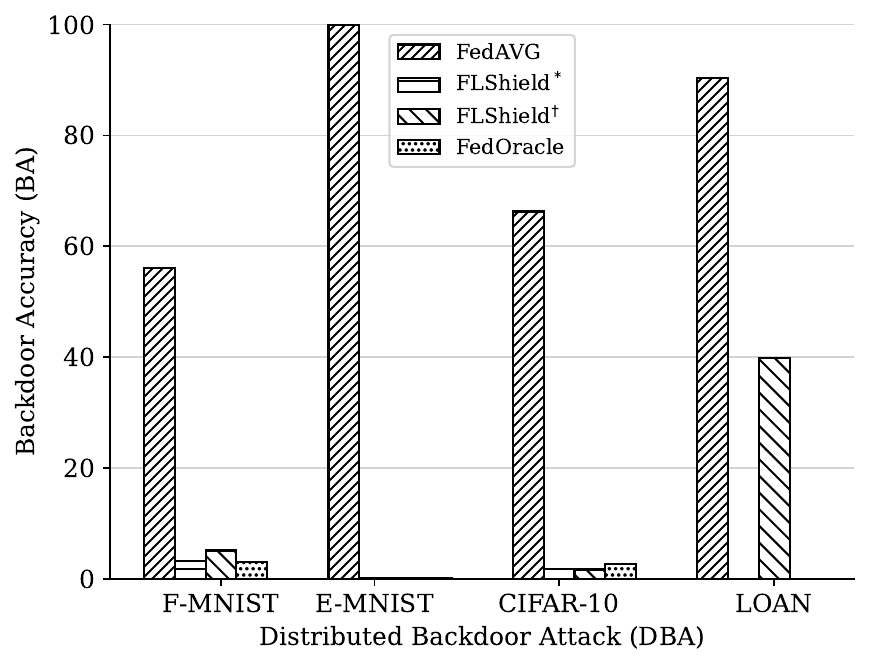}
        \label{fig:graph2_dba}
    }
    \hspace{-8pt}
    \caption{The performance of our defense against three different categories of poisoning attacks.}
    \label{fig:graph1}
\end{figure*}

\begin{table*}[t]
    \centering
    \tiny
    \caption{Effectiveness of \nameofdefense{} in comparison with state-of-the-art defenses against targeted poisoning attacks.}
    \label{tab:ours_vs_all_tlf}
    \begin{tabular}{llrrrlrrrlrrrlrrr}
         & \multicolumn{13}{c}{\textbf{Targeted Label Flipping Attack (TLFA)}} \\
         \hline
        \multirow{2}{*}{Defense}  &  \multicolumn{3}{c}{{F-MNIST}} & & \multicolumn{3}{c}{{E-MNIST}} &  & \multicolumn{3}{c}{{CIFAR-10}} & & \multicolumn{3}{c}{{LOAN}} \\
        \cline{2-16}
            &   RCL  &   ASR  &   MA  &    &   RCL  &   ASR  &   MA  &    &   RCL  &   ASR  &   MA &  & RCL  &   ASR  &   MA  \\
        \hline
        \texttt{FedOracle}  &           {81.97} &         6.4  &       87.76 &          &           99.26 &         0.2  &       99.39 &          &          83.60 &        4.46 &      78.76 & &95.93	&3.98	&95.42\\
        FedAvg       &           36.93 &        35.97 &       82.87 &          &           64.36 &        34.46 &       95.95 &          &          50.65 &       42.08 &      76.16 & &47.64	&52.13	&76.91\\
          RFA &           26.94 &        44.73 &       82.03 &          &           77.84 &        20.86 &       97.31 &          &          52.8    &        7.83 &      79.35 &&32.85	&65.74	&70.65\\
          AFA         &           36.91 &        35.97 &       82.88 &          &           58.09 &        40.76 &       95.29 &          &          48.64 &       44.77 &      75.79 & &37.84	&59.96	&72.88\\
          FLAME       &           42.04 &        41.4  &       81.64 &          &           78.2  &        20.68 &       97.26 &          &          48.92 &       45.38 &      72.88 & &82.18	&6.19	&90.16\\
          FLTrust     &           59.69 &        20.95 &       83.95 &          &           98.65 &         0.69 &       99.32 &          &          57.2  &       35.26 &      73.77 & &89.14	&7.22	&92.82\\
          \nameofdefense{}$^*$   &           \bf{84}    &         \bf{5.43} &       87.05 &          &           \bf{99.28} &         \bf{0.17} &       99.29 &          &          {81.10} &        \bf{5.27} &      \bf{78.96} & &97.37	 &2.56	&95.41\\
          \nameofdefense{}$^{\dagger}$   &           80.95 &         6.4  &       \bf{87.08} &          &           99.11 &         0.27 &       \bf{99.35} &          &          \bf{83.60} &        6.14 &      78.66 & &97.7	&2.22	&95.29\\
        \hline
        \end{tabular}
    \hfill
        \begin{tabular}{llrrlrrlrr}
                & \multicolumn{5}{c}{\textbf{Distributed Backdoor Attack (DBA)}} \\
                \hline
            \multirow{2}{*}{Defense}  &  \multicolumn{2}{c}{{F-MNIST}} &   & \multicolumn{2}{c}{{CIFAR-10}} \\
            \cline{2-6}
                 &   BA  &   MA  &    &   BA  &   MA  \\
            \hline
            \texttt{FedOracle}  &        3.03 &       87.72 &             &       2.75 &      79.06 \\
            FedAvg       &       56.15 &       87.07 &          &      66.27 &      78.94 \\
              RFA &       91.44 &       86.84 &          &       6.06 &      79.75 \\
              AFA         &       53.8  &       87.05 &                &      68.62 &      78.82 \\
              FLAME       &       6.17 &       86.36 &         &      4.68 &      78.81 \\
              FLTrust     &        6.5  &       86.99 &         &      20.15 &      77.45 \\
              \nameofdefense{}$^*$   &        \bf{3.25} &       \bf{87.64} &         &       {1.73} &      \bf{79.75} \\
              \nameofdefense{}$^{\dagger}$   &        5.13 &       87.18 &                &      \bf{1.69}    &      77.85 \\
            \hline
            \end{tabular}
\end{table*}

\section{Evaluation}

\subsection{Experiment Setup}
\label{subsec:expsetup}

\textbf{Datasets.}
We conduct experiments with four datasets: F-MNIST \cite{dataset_Fashion_mnist}, E-MNIST \cite{dataset_Extending_MNIST}, CIFAR-10 \cite{dataset_CIFAR}, and LOAN \cite{dataset_LOAN}. 
The first three pertain to the image domain while the last one represents tabular data. 
More details of the datasets are presented in Appendix~\ref{app:datasets}.

\vspace{0.1cm}
\noindent\textbf{Attacks considered.}
We evaluate \nameofdefense{} against three categories of attacks: poisoning attacks, privacy inference attacks, and attacks that are specifically designed against \nameofdefense{}.
The poisoning attacks include both untargeted (i.e., inner product manipulation attack or IPMA~\cite{fl_attack_Inner_Product_Manipulation}) and targeted (i.e., targeted label flipping attack or TLFA~\cite{tolpegin2020data} and backdoor) ones.
Further, we consider three backdoor attacks: distributed backdoor attack (DBA) \cite{dba}, edge-case backdoor attack (ECBA) \cite{wang2020attack}, and semantic backdoor attack (SBA) \cite{Bag2018}.
We also consider a gradient inversion attack (GIA)~\cite{gradient_inversion_2020} to evaluate if the representative models used in \nameofdefense{} can be used by the malicious validators to reconstruct training data from other clients.
As mentioned earlier, we design \nameofdefense{}-aware attacks where malicious participants attempt to compromise the integrity of the validation process. 
We design two such attacks: FLShield-aware adaptive attack (FA-Adp) and FLShield-aware advanced attack (FA-Adv).
Table \ref{tab:attack_taxonomy} (in Appendix) summarizes the attacks we investigate in our experiments. Section~\ref{app:attacks} provides more details of the attacks.

\vspace{0.1cm}
\noindent\textbf{Baselines.} We evaluate \nameofdefense{} against two baselines:
\begin{enumerate*}[label={(\arabic*)}]
\item FL aggregation without any defense (FedAvg), and
\item FL aggregation with perfect detection and removal of poisoned updates (\texttt{FedOracle}).
\end{enumerate*}
FedAvg's performance illustrates undefended attack effectiveness while \texttt{FedOracle} serves as a target baseline for any robust defense against FL poisoning.

\vspace{0.1cm}
\noindent\textbf{Existing defenses considered.} 
We compare \nameofdefense{}'s performance with the following state-of-the-art defenses: FLAME \cite{Flame}, FLTrust \cite{fltrust}, Adaptive Federated Averaging (AFA) \cite{AFA}, and Robust Federated Aggregation (RFA) \cite{rfa}.

\vspace{0.1cm}
\noindent\textbf{Metrics.} 
We consider the metrics main accuracy (MA), recall (RCL), and backdoor accuracy (BA) as defined earlier in section \ref{sec:preliminaries}.
These metrics evaluate the performance of the final global model.
To evaluate the effectiveness of \nameofdefense{} in filtering malicious model updates, we use true positive rate (TPR) and true negative rate (TNR).
TPR (TNR) is computed as the ratio of malicious (benign) model updates detected
as malicious (benign) and is averaged across all iterations.

\vspace{0.1cm}
\noindent\textbf{Data distribution.}
We consider both IID and non-IID scenarios.
For non-IID, we implement two strategies: one-class-expert distribution (also used in \cite{fltrust})
and Dirichlet distribution \cite{minka2000}.
In the first strategy, the clients are split into $10$ groups each corresponding to a class, and then each client is given an equal number of samples where $50\%$ of the samples are from the class corresponding to the client and the rest $50\%$ of the samples are from the other classes.
For the Dirichlet distribution \cite{minka2000}, we use the same setup as \cite{Bag2018}.

\vspace{0.1cm}
\noindent\textbf{FL setup.} Unless otherwise specified, we assume $40\%$ of the clients are malicious. We show the impact of varying the percentage of malicious clients (upto $45\%$) on \nameofdefense{}'s performance in section \ref{subsec:impact_of_malicious_clients}. The settings/configurations of the FL systems, classifier architecture, and hyperparameters of the attacks are described in Appendix section \ref{sec:appendix}.

\vspace{0.1cm}
\noindent\textbf{Code.} We used implementation from \cite{wang2020attack} and \cite{gradAttack} and extended it to cover the diverse range of attacks and defenses that we experimented with. Should this manuscript be accepted, we plan to make the associated code publicly available.

\subsection{Performance of \nameofdefense{}}
\label{sec:overall_eval}
We show the overall performance of \nameofdefense{} against IPMA, TLFA, and DBA in Figure \ref{fig:graph1} and the performance against EBCA and SBA are reported in Table \ref{tab:edge_case}\textcolor{green}{(b)}.
Both \nameofdefense{}$^{*}$ and \nameofdefense{}$^{\dagger}$ are able to achieve performances close to \texttt{FedOracle} for all datasets
against IPMA and TLFA.
While \nameofdefense{}$^{\dagger}$'s performance degrades when experimented with DBA on the LOAN dataset, \nameofdefense{}$^{*}$ is able to achieve a performance similar to \texttt{FedOracle} for all cases.
In section \ref{ablation:bijective} and section \ref{ablation:validation}, we conduct an in-depth analysis to show how the use of representative models and LIPC score contributes to \nameofdefense{}'s success.
In section \ref{ablation:backdoor}, we perform an ablation study to understand how \nameofdefense{} succeeds against backdoor attacks.
Also, we discuss the limitations of \nameofdefense{} in section~\ref{sec:limitation}.
According to our observation, \nameofdefense{} outperforms \texttt{FedOracle} in multiple instances. This is due to the fact that \nameofdefense{}'s design allows it to filter not only the malicious updates but also the updates that do not generalize well. Hence, \nameofdefense{}  produces global models that are often more generalized compared to that of \texttt{FedOracle} as \texttt{FedOracle} filters only the malicious updates.

\subsubsection{Comparison with Existing Defenses}
Table~\ref{tab:ours_vs_all_tlf} compares our proposed defenses with existing defenses.
The results show that both \nameofdefense{}$^{*}$ and \nameofdefense{}$^{\dagger}$ significantly outperform  the existing defenses.
Note that, both TLFA and DBA are covert attacks that do not significantly affect the main accuracy (MA) of the model.
State-of-the-art defenses such as FLTrust \cite{fltrust}, RFA \cite{rfa}, AFA \cite{AFA} fail to detect these attacks because these defenses have primarily been designed to defend against untargeted poisoning attacks only (details of IPMA results in Table~\ref{tab:ipma_result} in Appendix).
Even though FLAME~\cite{Flame} has been designed to detect targeted attacks (more specifically, backdoor attacks), it does not perform well against TLFA.
This is because HDBSCAN is not able to separate the clusters of the benign and malicious clients in the TLFA setting.
In contrast, \nameofdefense{}$^{*}$ and \nameofdefense{}$^{\dagger}$ are able to achieve performances close to \texttt{FedOracle} in all datasets against both TLFA and DBA across all metrics.

\begin{figure}[t]
    \centering
    \subfigure[]{
        \includegraphics[width=0.49\linewidth]{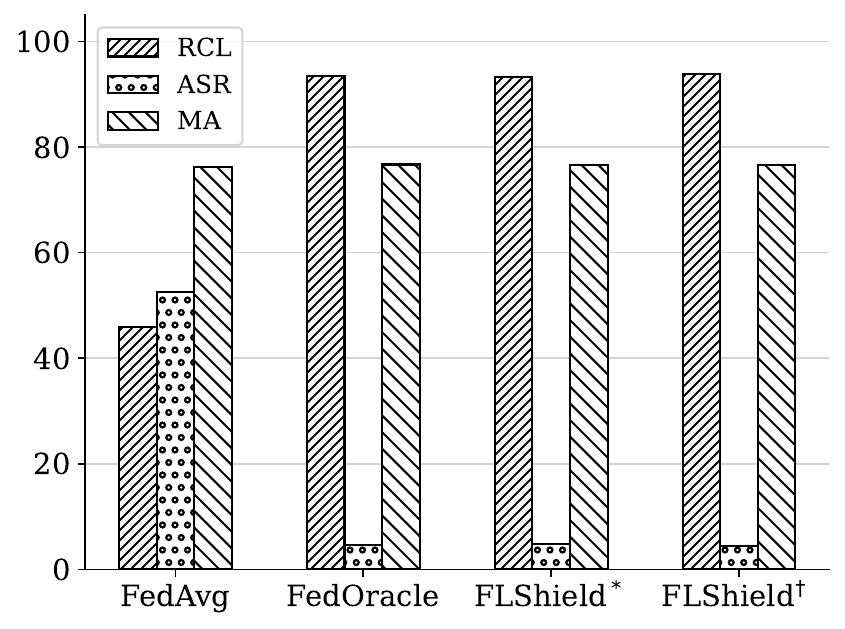}
    \label{fig:noniid_one_class_expert}
    }
    \hspace{-22pt}
    \subfigure[]{
        \includegraphics[width=0.49\linewidth]{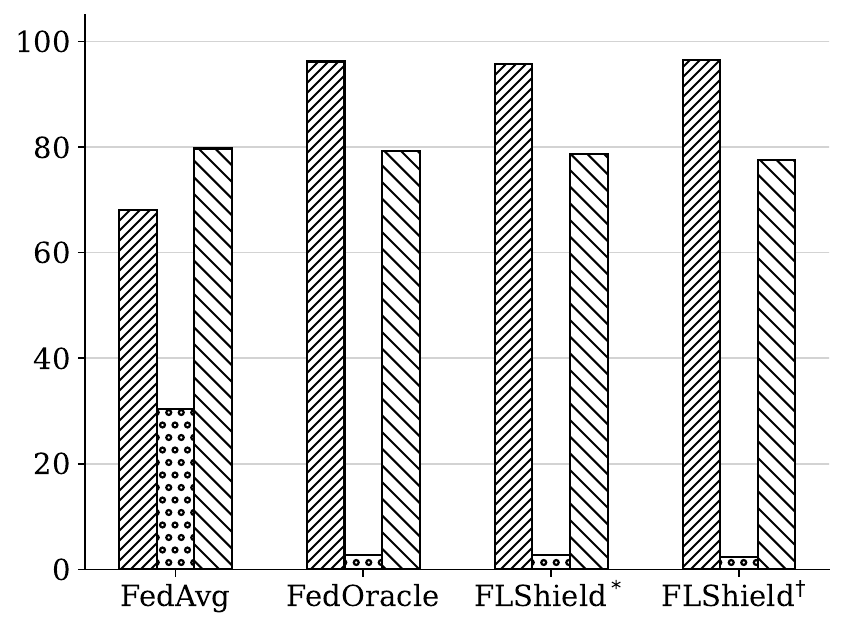}
\label{fig:FLTrust_fail}
    }
    \caption{Evaluation of \nameofdefense{} in non-IID scenario: (left) one-class-expert, (right) Dirichlet}
    \label{fig:noniid}
\end{figure}

\subsubsection{Performance in Non-IID Settings}
\label{subsec:non_iid}

We experiment with two non-IID strategies as mentioned in section~\ref{subsec:expsetup}. Also, note that, the LOAN dataset has a natural non-IID distribution. 
Figure \ref{fig:graph1} shows that \nameofdefense{} is able to achieve performances comparable to \texttt{FedOracle} for the LOAN dataset.
Figure \ref{fig:noniid} shows the performance against TLFA attack on F-MNIST dataset in non-IID settings. For both one-class-expert and Dirichlet distributions, \nameofdefense{} performs similar to \texttt{FedOracle}.
This is interesting since the validators are picked from the non-IID client groups as well.
More specifically, this result shows that the validation mechanism of \nameofdefense{} does not require the validators to be IID. 
We deem the following as the primary reason for \nameofdefense{}'s success in non-IID scenarios.

\begin{figure}
    \centering
    \subfigure[With original $\mathcal{N}$]{
        \includegraphics[width=0.47\linewidth]{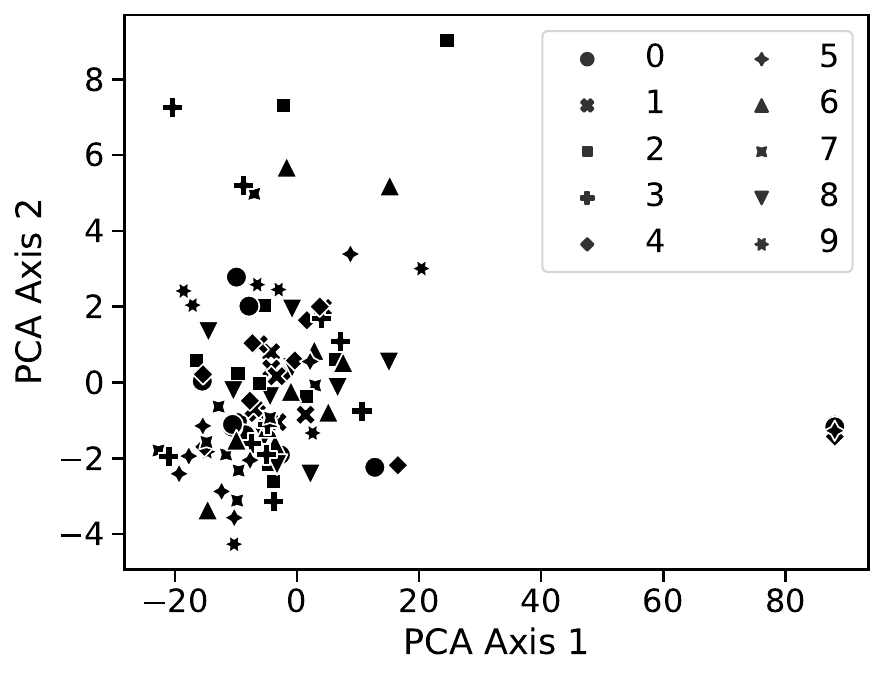}
        \label{fig:noniid_actual_pca}
    }
    \hspace{-12pt}
    \subfigure[With modified $\mathcal{N}$]{
        \includegraphics[width=0.49\linewidth]{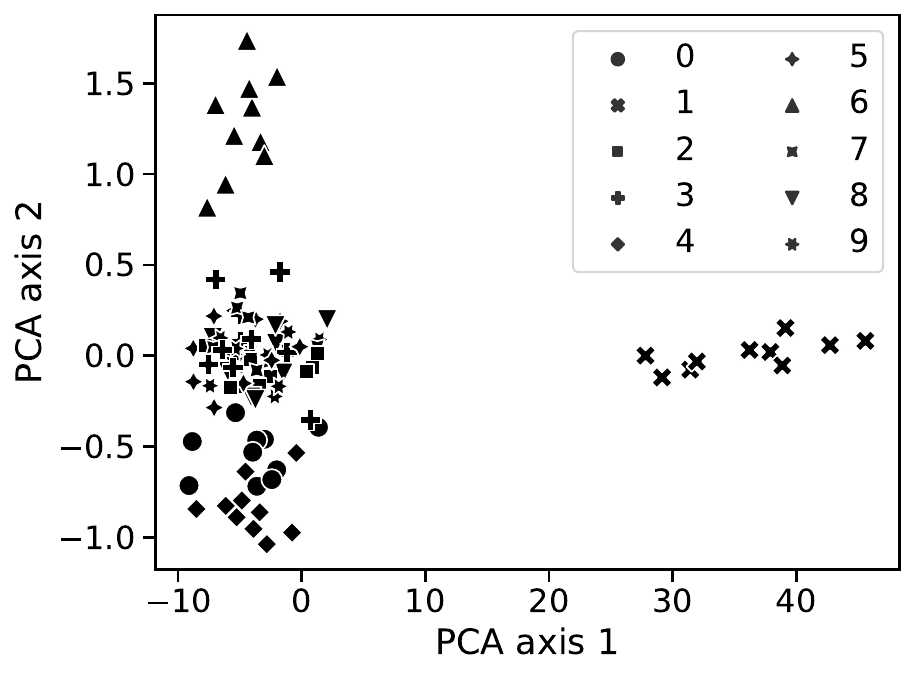}
        \label{fig:noniid_alternate_pca}
    }
    \caption{
    PCA projection of $\mathcal{N}$ (left) and a modified version of $\mathcal{N}$ (right) of 100 validators in the one-class-expert setting.
    Each validator belongs to one of the 10 groups denoted by the class index of the dominant class in their training data.
    }
\end{figure}

\begin{itemize} [leftmargin=0pt, itemindent=15pt]
\item \emph{The $\mathcal{N}$ matrices of validators are independent from validation data distribution.} We flattened each $\mathcal{N}$ matrix to a vector and projected the vectors after performing PCA in Figure \ref{fig:noniid_actual_pca}. The $\mathcal{N}$ matrices are taken from a one-class-expert setting with the F-MNIST dataset and as such there are 10 groups of clients with each group defined by the dominant class sample in clients' validation data. The figure shows that the $\mathcal{N}$ matrices of validators are similar irrespective of the group they belong to. In fact, the intra-group mean distance of $\mathcal{N}$ matrices is 0.0126 which is close to the inter-group mean distance of $\mathcal{N}$ matrices of 0.0128. Hence, it is evident that the $\mathcal{N}$ matrix is not influenced by the validation data distribution. This is because the $\mathcal{N}$ matrix is computed in a class-wise manner. For a particular class, a validator from the corresponding group has more samples of the class than a validator from another group. If the $\mathcal{N}$ was computed as only a vector containing the loss impact of all the representative models on the validators' data, then it would be biased by the dominant class of the validator. Figure \ref{fig:noniid_alternate_pca} shows the PCA projection of $\mathcal{N}$ vectors in this alternate scenario. Some groups of validators are reporting similar results to a group member but disparate results from a non-member. The similarity of original $\mathcal{N}$ matrices implies the absence of disagreement among the validators which in turn enables \nameofdefense{} to defend well in non-IID scenarios.
\end{itemize}

\subsection{\nameofdefense{}-aware Attacks by Malicious Validators}
\label{sec:mal_val_attack}

To evaluate \nameofdefense{}'s robustness, we assume an enhanced adversary which has the knowledge of \nameofdefense{} and has additional capabilities to:
\begin{enumerate*}[label={(\arabic*)}]
    \item manipulate the $\mathcal{N}$ matrices computed by malicious validators, and
    \item estimate the $\mathcal{N}$ matrices computed by benign validators
\end{enumerate*}.

The enhanced adversary aims to achieve two objectives:
\begin{enumerate*}[label={(\arabic*)}]
    \item \emph{Stealth objective}: ensures that the $\mathcal{N}$ matrices by the malicious validators cannot be distinguished from that of the benign validators by FLShield's outlier detection algorithm, and
    \item \emph{Infiltration objective}: ensures that the malicious representative models rank in the top $50\%$.
\end{enumerate*}
However, it is important to note that finding a set of crafted validation results that meet both objectives could be difficult for the adversary. This is because, for example, malicious representative models able to poison the FL system may not be stealthy.
If these objectives cannot be achieved simultaneously, we can conclude that any attack that is aware of \nameofdefense{} would still fail to poison the FL system. We design two enhanced adversaries to meet these two objectives at the same time:

\begin{enumerate}[label=\textbf{(\arabic*)},  leftmargin=0pt, itemindent=15pt]
\item \textbf{FLShield-aware adaptive attack (FA-Adp):}
In this approach, the adversary tries to achieve the two objectives through the formulation of an optimization problem. Such adaptive attacks have been designed to prove the effectiveness of other defenses in the literature, e.g., in FLTrust~\cite{fltrust}.
In this tactic, we define a loss function combining the two objectives. 
The optimal $\mathcal{N}$ matrices are then derived by utilizing the stochastic gradient descent method to minimize the loss function.
The strategy details are presented in Appendix \ref{appendix:mal_val_attack} due to space constraints.

\item \textbf{FLShield-aware advanced attack (FA-Adv):}
In this approach, the malicious validators first compute the true $\mathcal{N}$ matrices and then attempt to achieve their stealth and infiltration objectives by making minimal alterations to these original validation results. 
The magnitude of alteration can be quantified as the distance between the original and the crafted $\mathcal{N}$ matrix by each malicious validator. 
While finding the minimal change necessary to meet the infiltration objectives is mathematically intractable, a suboptimal solution can be derived by following the steps detailed in the Algorithm presented in Figure~\ref{alg:naive} in Appendix.

\end{enumerate}

\emph{Evaluation Results:}
Table \ref{tab:mal_val_attack} compares the performance of \nameofdefense{} against FA-Adp and  FA-Adv attacks including the case when there is no malicious validator.
In the no malicious validator scenario, when a malicious participant is selected as a validator it does not manipulate the validation results. 
The attacks in this experiment are performed on the CIFAR-10 dataset.
The results show that both versions of \nameofdefense{}-aware attacks achieve limited success in terms of RCL, BA, TPR, and TNR. The only anomaly is the performance of \nameofdefense{}$^{*}$ against the FA-Adv attack that reduces TPR and TNR by ~10\% under DBA.
Figures \ref{fig:lipc}, \ref{fig:fa-lipc} display representative models ranked by $\mathcal{L}$ score under DBA and TLFA respectively. 
DBA exhibits a smaller benign-malicious model gap compared to TLFA. 
FA-Adv's strategy, aiming to narrow this gap, only finds limited success when minimum alterations don't yield out-of-distribution $\mathcal{N}$. 
However, such occurrences are rare, and the limited infiltration is inadequate for an effective backdoor.
\bluesp{
The recall is higher in some defense-aware attack scenarios compared to non-defense-aware ones.
However, recall is not a robust metric for evaluating the defense's effectiveness due to various influencing factors in an FL system. 
Differences in benign model updates selected in cases like None and FA-Adp or FA-Adv lead to different recall values, even if they are close. 
TPR/TNR, in contrast, serves as a direct measure of the defense's ability to eliminate poisoned model updates.
}
In section \ref{ablation:validation}, we perform an ablation study to closely understand the reason behind the success of \nameofdefense{} against malicious validator attacks.

\begin{table}[]
\centering
\scriptsize
\resizebox{0.99\columnwidth}{!}{
\begin{tabular}{llrrrrrr}
\hline
                    \multicolumn{2}{c}{Malicious Validators $\rightarrow$}                                                          & {None}  & {FA-Adp} & FA-Adv & {None}  & {FA-Adp} & FA-Adv \\ \hline 
                    Metrics $\downarrow$ &    Attacks $\downarrow$                                                            & \multicolumn{3}{c}{\nameofdefense{}$^*$}                                      & \multicolumn{3}{c}{\nameofdefense{}$^{\dagger}$ }                                      \\ \hline
RCL  & TLFA & {81.10} & {83.10} & 81.60    & {83.60} & {84.80} & 83.70       \\ \hline
                   BA  & DBA                                                          & {1.73}  & {1.72}  & 2.47     & {1.69}  & {1.48}  & 1.87     \\ \hline
\multirow{2}{*}{TPR} & TLFA & {100}   & {100}   & 100      & {100}   & {100}   & 100    \\ 
                     & DBA                                                            & {100}   & {100}    & 90      & {100}   & {100}   & 100    \\ \hline
\multirow{2}{*}{TNR} & TLFA & {91}    & {90.75} & 90.75    & {86.67} & {86.67} & 86.67      \\ 
                     & DBA                                                            & {100}   & {100}    & 90      & {86.67} & {86.67} & 86.67      \\ \hline
\end{tabular}
}
\caption{Evaluation of \nameofdefense{} under FA-Adp and FA-Adv}
\label{tab:mal_val_attack}
\end{table}

\ignore{
\begin{table}[]
\scriptsize
    \centering
    \begin{tabular}{llrrr}
\toprule
{} & Performance Metric &  Adaptive &  Naive &   None \\
\midrule
TLF &             Recall &     90.08 &  87.99 &  89.07 \\
DBA &  Backdoor Accuracy &     15.17 &  24.56 &   9.86 \\
IPM &      Main Accuracy &     78.47 &  78.67 &  78.07 \\
\bottomrule
\end{tabular}
    \caption{Performance comparison of different malicious validator attacks}
    \label{tab:table_mal_val_1}
    
\end{table}

\begin{table}[]
\scriptsize
    \centering
    \begin{tabular}{lrrr}
\toprule
{} &  Adaptive &  Naive &  None \\
\midrule
TLF &      0.00 &   0.00 &  0.00 \\
DBA &      0.05 &   0.07 &  0.04 \\
IPM &      0.05 &   0.00 &  0.03 \\
\bottomrule
\end{tabular}
\caption{Percentage of malicious representatives that got accepted in the malicious validator attack experiments}
\label{tab:table_mal_val_2}
\end{table}
}

\begin{figure}
    \centering
    \includegraphics[width=0.9\linewidth]{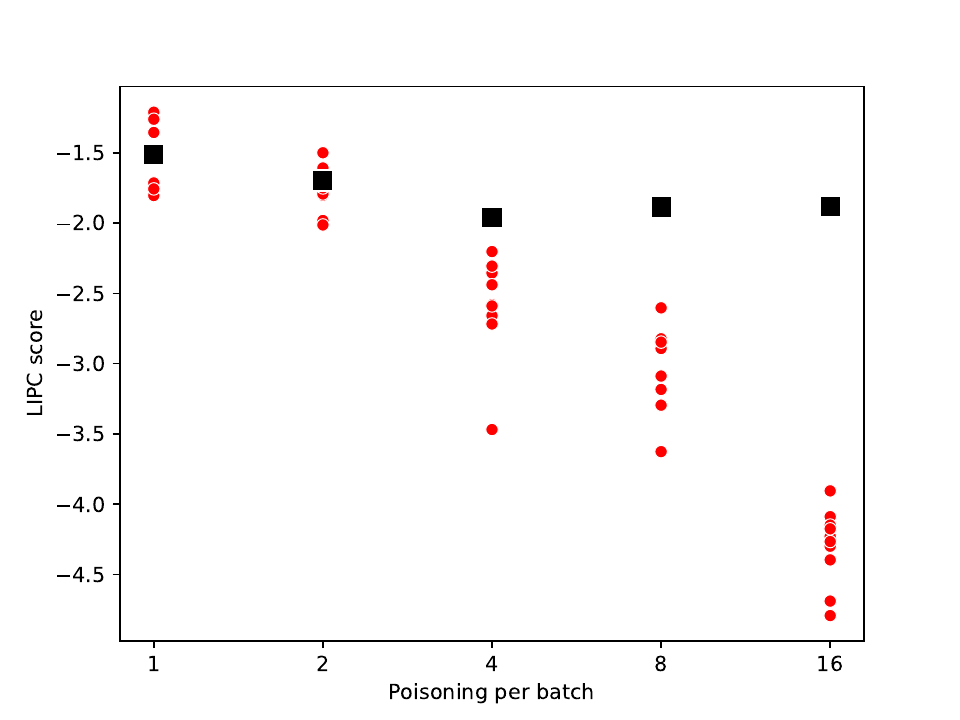}
    \caption{
    LIPC score of malicious representative models vs poisoned sample per batch (PSPB) in DBA. The square dot indicates the median LIPC score for the corresponding setting.
    }
    \label{fig:ppb_vs_lipc}
\end{figure}

\begin{figure*}[t]
  \centering
  \subfigure[IID]{
        \includegraphics[width=0.21\textwidth]{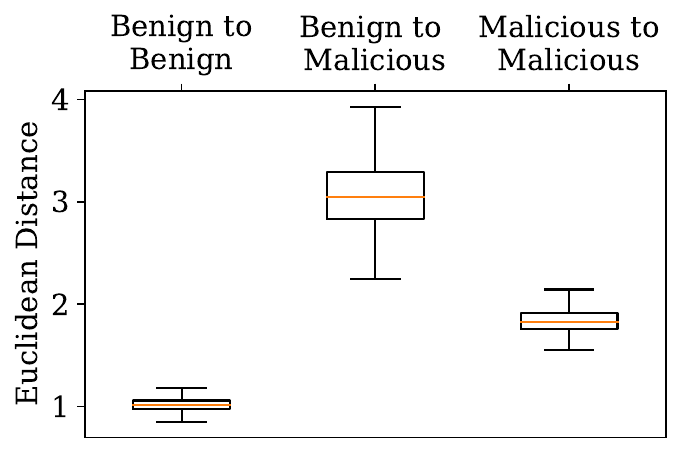}
        \label{fig:ablation_similarity_report_1}
  }
  \hspace{-6pt}
  \subfigure[One-class-expert]{
        \includegraphics[width=0.21\textwidth]{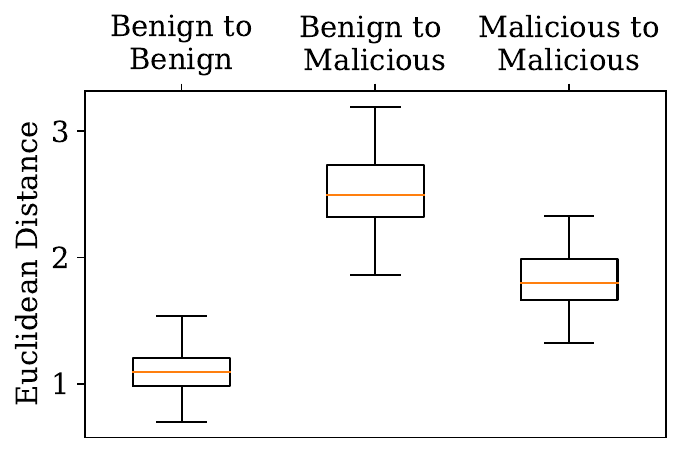}
        \label{fig:ablation_similarity_report_2}
  }
  \hspace{-6pt}
  \subfigure[Dirichlet]{
        \includegraphics[width=0.21\textwidth]{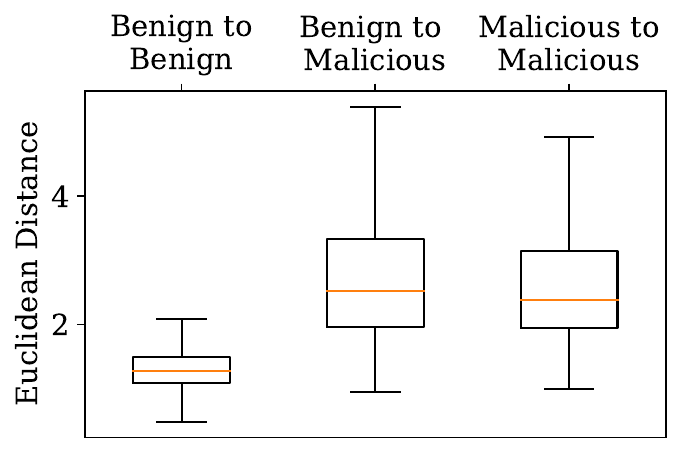}
        \label{fig:ablation_similarity_report_3}
  }
  \hspace{-6pt}
  \subfigure[\parbox{2cm}{\centering Benign\\ Representative \\Model}]{
        \includegraphics[width=0.145\textwidth]{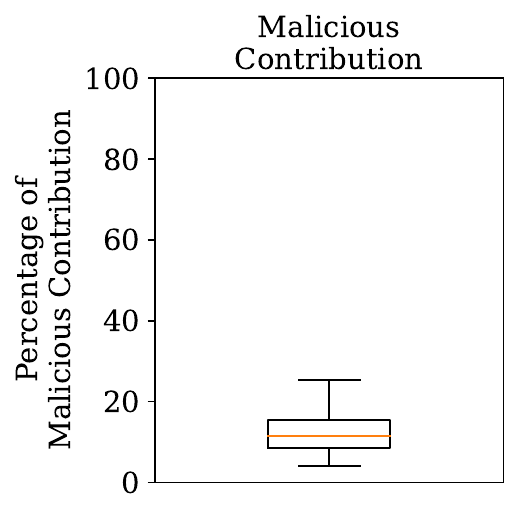}
    \label{fig:ablation_contribution_report4}
  }
\hspace{-6pt}
  \subfigure[\parbox{2cm}{\centering Malicious\\ Representative\\ Model}]{
        \includegraphics[width=0.15\textwidth]{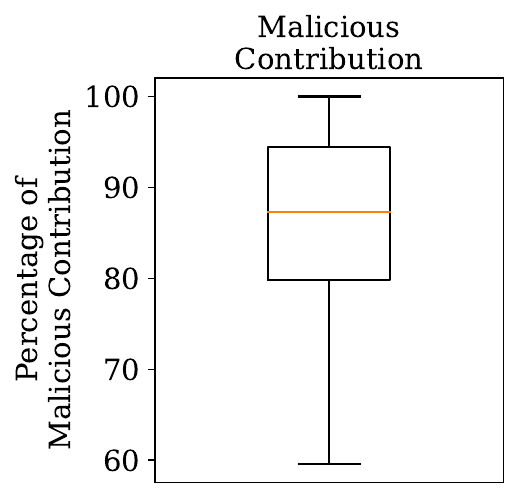}
    \label{fig:ablation_contribution_report5}
  }
  \caption{
  Euclidean distance between updates for three distribution settings: (a) IID, (b) one-class-expert, and (c) Dirichlet. Percentage of malicious contribution to (d) benign representative models and (e) malicious representative models.}
  \label{fig:boxplots}
\end{figure*}

\subsection{Ablation Study}

\subsubsection{Why does \nameofdefense{} work?}
\label{ablation:bijective}
\begin{figure*}
\centering
    \subfigure[]{
        \includegraphics[width=0.46\columnwidth]{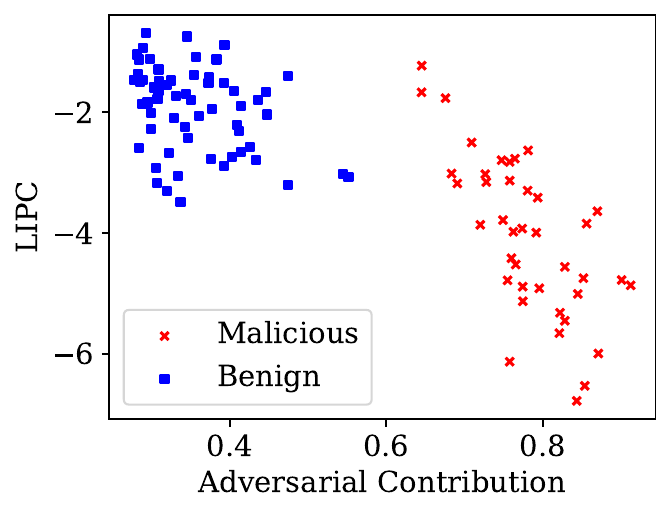}
        \label{fig:adv_contrib_vs_score}
    }
    \hspace{-12pt}
    \subfigure[]{
    \includegraphics[width=0.41\columnwidth, height = 3cm]
{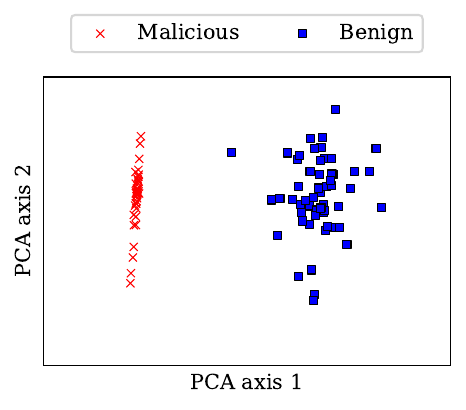}
    \label{fig:pca_validator}
    }
    \hspace{-17pt}
    \subfigure[]{
    \includegraphics[width= 0.5\columnwidth, height=3.3 cm]{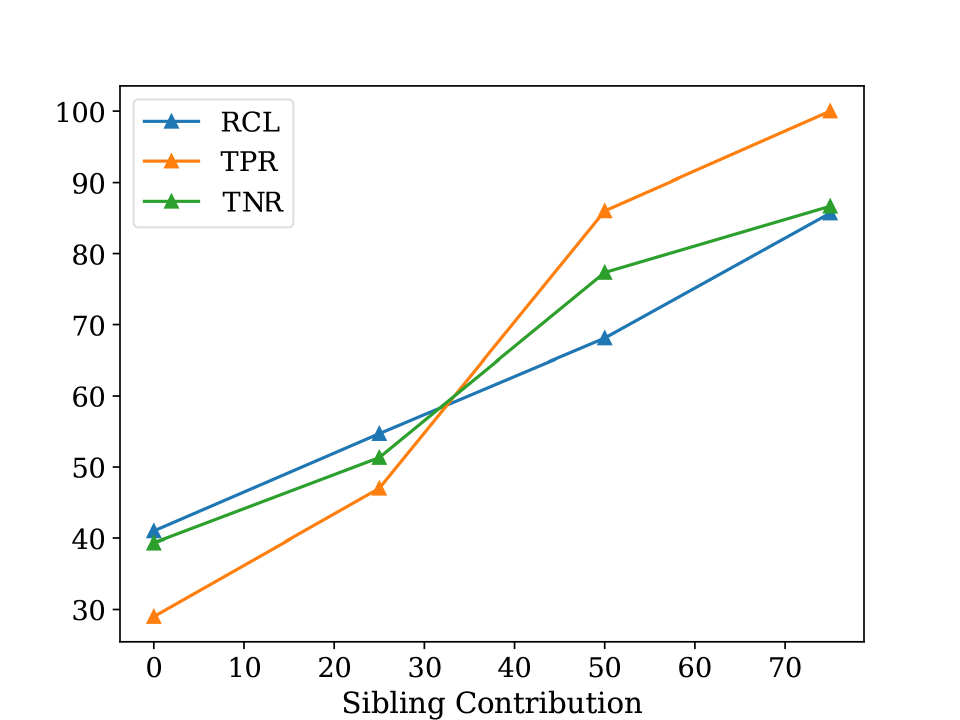}
    \label{fig:contrib_adj_1}
    }
    \hspace{-22pt}
    \subfigure[]{
    \includegraphics[width=0.25\columnwidth, height=3.3 cm] {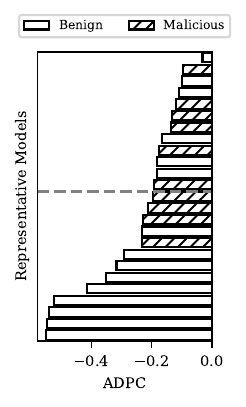}
    \label{fig:accuracy}
    }
    \hspace{-20pt}
    \subfigure[]{
    \includegraphics[width=0.25\columnwidth, height=3.3 cm] {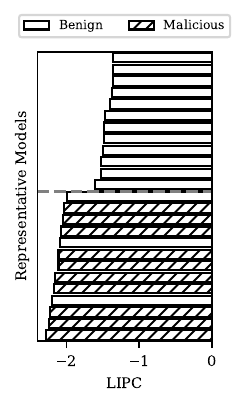}
    \label{fig:lipc}
    }
    \hspace{-19pt}
    \subfigure[]{
    \includegraphics[width=0.25\columnwidth, height=3.3 cm] {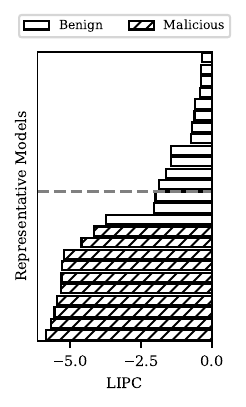}
    \label{fig:fa-lipc}
    }
    \caption{(a) LIPC vs. adversarial contribution,  (b) visualization of validation reports after PCA projection, (c) performance with varying sibling contributions, and (d-f) representative models ranked by metrics: (left) ADPC, (middle) LIPC under DBA, (right) LIPC under TLFA.}
\end{figure*}

The following list of observations coupled with visualizations from experimental results shed light on the reasons behind the success of \nameofdefense{}.
\begin{itemize} [leftmargin=0pt, itemindent=15pt]
    \item \emph{The  similarity between two benign (malicious) updates is greater than the  similarity between a benign update and a malicious update}. Models trained with similar objectives are expected to bear resemblance with each other. Figure \ref{fig:ablation_similarity_report_1}, \ref{fig:ablation_similarity_report_2}, and \ref{fig:ablation_similarity_report_3} present results from the experiments to support this. The three boxplots show the range of distances between two benign updates, benign and malicious updates, and two malicious updates in three distribution settings: IID, one-class-expert, and Dirichlet. In all cases, the mean benign-benign distance is smaller than the mean malicious-benign distance. The experiments consider TLFA on F-MNIST dataset.
    
    \item \emph{The contribution of  benign (malicious) local models to a benign (malicious) representative model is greater than the contribution of the malicious (benign) local models}. This is because the contributions are weighted based on the contributor's similarity with the base model. Figure \ref{fig:ablation_contribution_report4} and \ref{fig:ablation_contribution_report5} show the contribution of malicious participants to the benign and malicious representative models, respectively. 
    
    \item \emph{The performance of a representative model declines with an increase in the ratio of the malicious updates' contribution to that model}. In other words, more malicious contribution implies worse performance in terms of $\mathcal{L}$ scores. 
    
    Figure \ref{fig:adv_contrib_vs_score} presents evidence of this behavior by showing the $\mathcal{L}$ scores of 100 bijective representative model updates. The scores reflect a single iteration of FL with CIFAR-10 dataset and Dirichlet distribution where the adversary is launching TLFA.
    
    \item \emph{All the representative models that have a malicious majority contribution rank at the bottom $50\%$ of all the representative models in terms of performance.} The decreasing $\mathcal{L}$ scores with increasing malicious contributions in Figure \ref{fig:adv_contrib_vs_score} complies with the above. 
\end{itemize}

\subsubsection{How does \nameofdefense{} differentiate malicious validation reports from benign ones?}
\label{ablation:validation}
We perform an ablation study on the validation components of \nameofdefense{} and conduct a comprehensive inspection to evaluate the efficacy of our filtering mechanisms. Our analysis helps us identify some key reasons behind \nameofdefense{}'s ability to filter malicious validation results.

\begin{itemize}[leftmargin=0pt, itemindent=15pt]

    \item \emph{The use of $\mathcal{L}$ as validation metric and the incorporation of sibling contribution into the representative models make the benign validation reports congruent and difficult to manipulate.} Figure \ref{fig:pca_validator} casts the validation reports into two dimensions using PCA and portrays the congruency among the benign validators and the anomalous nature of the tailored validation reports created by the malicious validators.  \nameofdefense{} is able to achieve $100\%$ TPR when sibling contribution rate $\tau$ is set to $75\%$, i.e., base model contributes the rest $25\%$. We use this as the default ratio for our experiments. The effects of varying sibling contributions in terms of RCL, TPR, and TNR are demonstrated in Figure \ref{fig:contrib_adj_1}. The results also show that when local models are validated directly, i.e., when $\tau$ is set to $0\%$, \nameofdefense{}'s defense fails. Sibling contributions' success can be attributed to the improved generalization ability brought to the representative models as described in section~\ref{subsec:representative_design}.
    
    \item \emph{Stealthy poisoning can be identified using $\mathcal{L}$ metric whereas accuracy-based metric fails.} To demonstrate the significance of $\mathcal{L}$ metric, we design another metric \emph{accuracy difference per class} (ADPC) and define it as the difference between accuracy values output by the global model and the representative model for each class. Next, we run experiments with an altered version of \nameofdefense{}$^{\dagger}$ that uses this new metric instead of $\mathcal{L}$. Figure \ref{fig:lipc} and \ref{fig:accuracy} presents the results from an experiment with DBA. The results show that malicious and benign representative models are indistinguishable when measured with ADPC. \ignore{We also present the results with default version of \nameofdefense{}$^{\dagger}$ using $\mathcal{L}$ in \ref{fig:lipc} and \ref{fig:fa-lipc}. In this case, the difference between malicious and benign representative models are more evident. The finding from this comparison is interesting since backdoor poisoning involves minimizing the main loss function in addition to the backdoor loss function. One potential explanation could be that minimizing the loss function does not necessarily propel it to minimize loss for each class. It would be interesting to see what happens when the backdoor training objective is modified to incorporate the latter case. We leave it for future work.} The experiment results shown in Table \ref{tab:lipc_vs_accuracy}\textcolor{green}{(b)} further ascertains that ADPC is not a suitable metric to use in \nameofdefense{}. The experiment has been  performed on CIFAR-10 dataset with \nameofdefense{}$^{\dagger}$ used in aggregation and the adversary launches TLFA.
    
    \item \emph{Due to the robustness of the $\mathcal{L}$ metric, the choice of outlier detector algorithm to filter malicious validation reports does not matter significantly.} We experimented with Elliptic envelope, Isolation forest, Local outlier factor and found no difference in terms of performance among them (section \ref{sec:outlier_detector_experiment}). Note that, not using an outlier detector in the presence of malicious validators  leads to failure which signifies the importance of outlier detection. The experimental results are presented in Table \ref{tab:ablation_no_detector}\textcolor{green}{(a)} for CIFAR-10 dataset where \nameofdefense{}$^{\dagger}$ defends against TLFA with malicious validators performing FA-Adv. 
\end{itemize}

\subsubsection{How does \nameofdefense{} defend against backdoor attacks?}
\label{ablation:backdoor}
We perform an ablation study on various parameters of the backdoor attacks to understand the reasons behind \nameofdefense{}'s success against backdoor attacks.
\begin{itemize}[leftmargin=0pt, itemindent=15pt]
    \item \emph{LIPC score of malicious representative models is negatively correlated with poisoned sample per batch (PSPB) of the backdoor attack.} We conducted experiments under DBA with varying numbers of PSPB and reported the LIPC score of the malicious representative models in Figure \ref{fig:ppb_vs_lipc}. It is evident that the more backdoored samples are introduced to the training the more the LIPC score decreases from the median value. Backdoor poisoning involves both clean samples and backdoored samples to reduce the loss on the main task and backdoored task respectively. The increase of backdoored samples means the decrease of clean samples which in turn results in diminished loss reduction on the main task and that is reflected in the poor LIPC score of the malicious representative models.
    \item \emph{The required PSPB to bypass \nameofdefense{} is ineffective at setting up backdoors in the global model.} In Figure \ref{fig:ppb_vs_lipc}, it is shown that the malicious representative models get ranked in the top 50\% with an extremely low poisoned sample per batch. However, if the PSPB is below 8, the poisoning is ineffective even in FedAvg i.e. the final global model does not perform well on the backdoor task (BA=3.31\% at PSPB 4). Thus, it is clear that any effective backdoor poisoning requires a certain level of PSPB which in turn results in a discriminative LIPC score of the malicious representative models enabling \nameofdefense{} to defend against backdoor attacks.
\end{itemize}

\begin{table*}
    \centering
    \tiny
      \begin{minipage}[t]{0.25\textwidth}
    \centering
    \begin{tabular}{lccc}
\toprule
 & RCL & TPR & TNR \\
\midrule
No Detector & \textbf{15.70} & \textbf{10.00} & 26.67 \\
Default & 83.40 & 100.00 & 86.67 \\
\bottomrule
\end{tabular}
\caption*{(a)}
  \end{minipage}
  \hspace{-10pt}
  \begin{minipage}[t]{0.25\textwidth}
    \centering
        \begin{tabular}{lcc}
\toprule
Metric &  \multirow{2}{*}{RCL} &  \% of Malicious \\  used& & Representatives Accepted \\
\midrule
ADPC &   25.34 &  26 \\
LIPC     &   88.66 &  0 \\
\bottomrule
\end{tabular}
\caption*{(b)}

  \end{minipage}
\hspace{-10pt}
\begin{minipage}[t]{0.25\textwidth}
    \centering
    \begin{tabular}{lrrr}
\toprule
 & RCL & TPR & TNR \\
\midrule
Mean & 73.00 & \textbf{80.60} & 73.73 \\
Minimum & 83.60 & 100.00 & 86.67 \\
\bottomrule
\end{tabular}
\caption*{(c)}
    
  \end{minipage}
  \hspace{-10pt}
  \begin{minipage}[t]{0.25\textwidth}
    \centering
    \begin{tabular}{lrrr}
\toprule
\# of Validators & BA & TPR & TNR \\
\midrule
10 & 3.61 & \textbf{97.50} & 81.67 \\
15 & 3.46 & 100.00 & 88.89 \\
20 & 2.64 & 100.00 & 83.33 \\
25 & 1.89 & 100.00 & 86.67 \\
\bottomrule
\end{tabular}
\caption*{(d)}
    
  \end{minipage}

  \caption{(a) Result Comparison between two scenarios: (1) no outlier detector (2) default version of \nameofdefense{}.(b) Performance comparison between two validation metrics: LIPC vs Accuracy-difference-per-class.(c) Performance comparison between mean and minimum (default) as the projection step in Algorithm \ref{alg:filtering} Line 4. (d) BA, TPR, and TPR vs. number of validators.}
  
  \label{tab:ablation_no_detector}
    \label{tab:lipc_vs_accuracy}
  \label{tab:ablation_use_mean}
  \label{tab:ablation_num_of_validators}
\end{table*}

\begin{table}[htbp]
\tiny
  \begin{minipage}{.56\linewidth}
    \centering
    \begin{tabular}{lllll}
\hline
Aggregation & Malicious & TPR & TNR & BA \\
of & contribution \% &&&\\
\hline
Representative & \textbf{8.60} & 99 & 86 & \textbf{51} \\
Individual & N/A & 100 & 86.67 & 0.43 \\
\hline
\end{tabular}
\caption*{(a)}

  \end{minipage}%
  \begin{minipage}{.5\linewidth}
    \centering
    \begin{tabular}{lcc}
\hline
\multirow{2}{*}{Defense} & ECBA  & SBA \\ \cline{2-3} 
                         & BA    & BA  \\ \hline
FedAvg                   & 85.71 & 100 \\ \hline
\texttt{FedOracle}                & 8.67  & 0   \\ \hline
\nameofdefense{}$^*$                & 9.18  & 0   \\ \hline
\nameofdefense{}$^{\dagger}$                & 7.65  & 0   \\ \hline
\end{tabular}
    \caption*{(b)}
  \end{minipage}
  \caption{(a) Aggregation of representative models vs. individual local models. Dataset: CIFAR-10, Attack: DBA. (b) Performance evaluation of  ECBA and SBA on CIFAR-10 }

  \label{tab:ablation_aggregate_ensemble}
  \label{tab:edge_case}
\end{table}

\begin{figure}[t]
    \centering
    \subfigure{
        \includegraphics[width=0.49\linewidth]{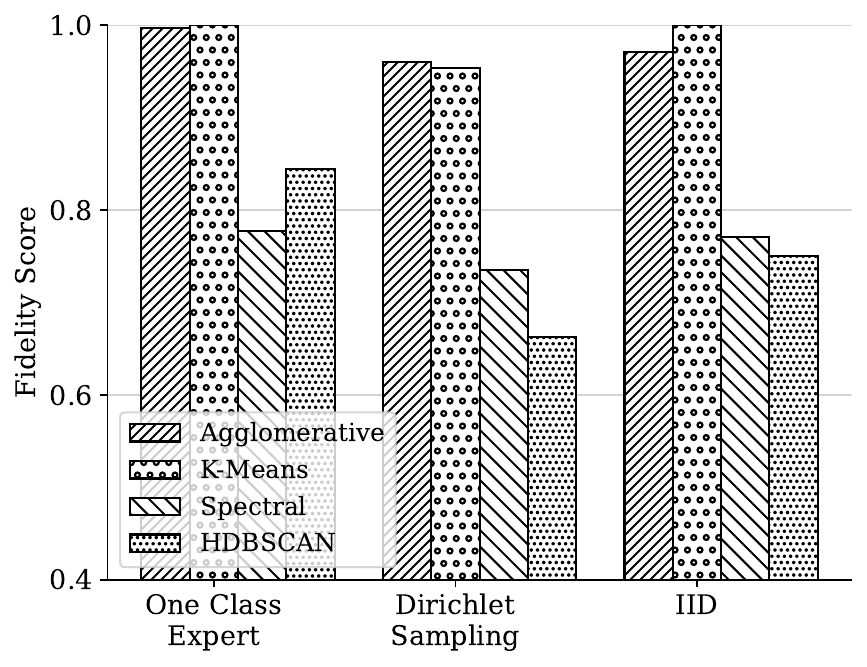}
        \label{fig:cluster_comparison}
    }
    \hspace{-16pt}
    \subfigure{
        \includegraphics[width=0.49\linewidth]{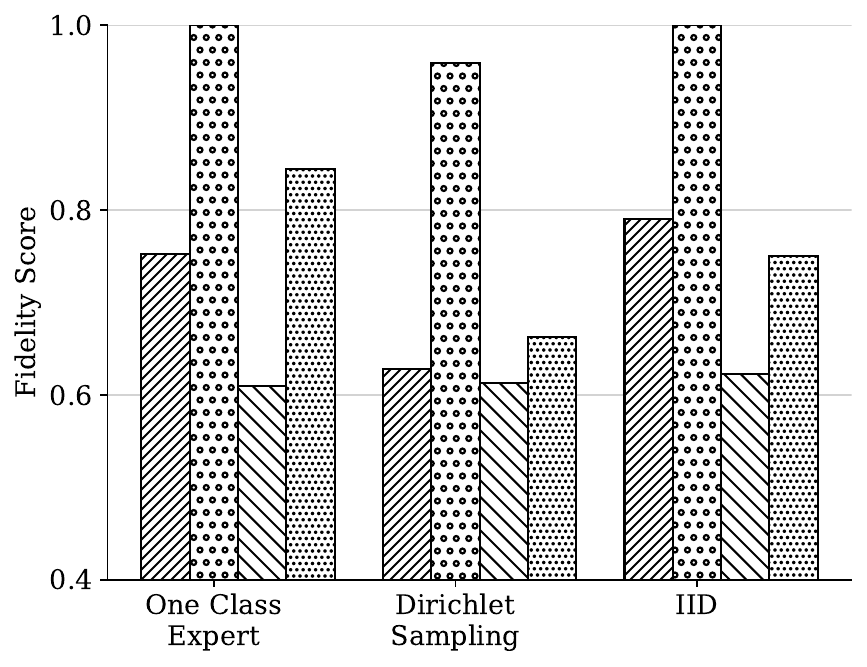}
        \label{fig:cluster_comparison_n_cluster_2}
    }
    \caption{Fidelity score comparison of multiple clustering algorithms with: (left) dynamic clustering, (right) fixed number of clusters (2). Dataset: F-MNIST, attack: TLFA.}
    \label{fig:ablation_clustering}
\end{figure}

\begin{figure*}[t]
    \centering
    \subfigure[]{
        \includegraphics[width=0.26\textwidth]{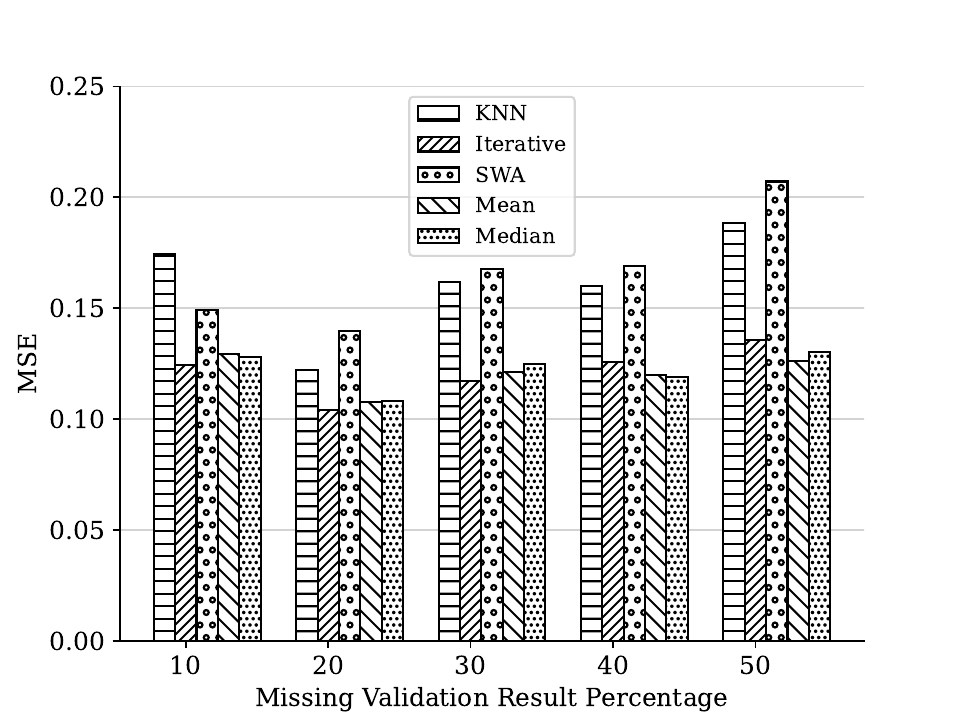}
        \label{fig:ablation_imputation_comparison_iid}
    }
    \hspace{-22pt}
    \subfigure[]{
        \includegraphics[width=0.26\textwidth]{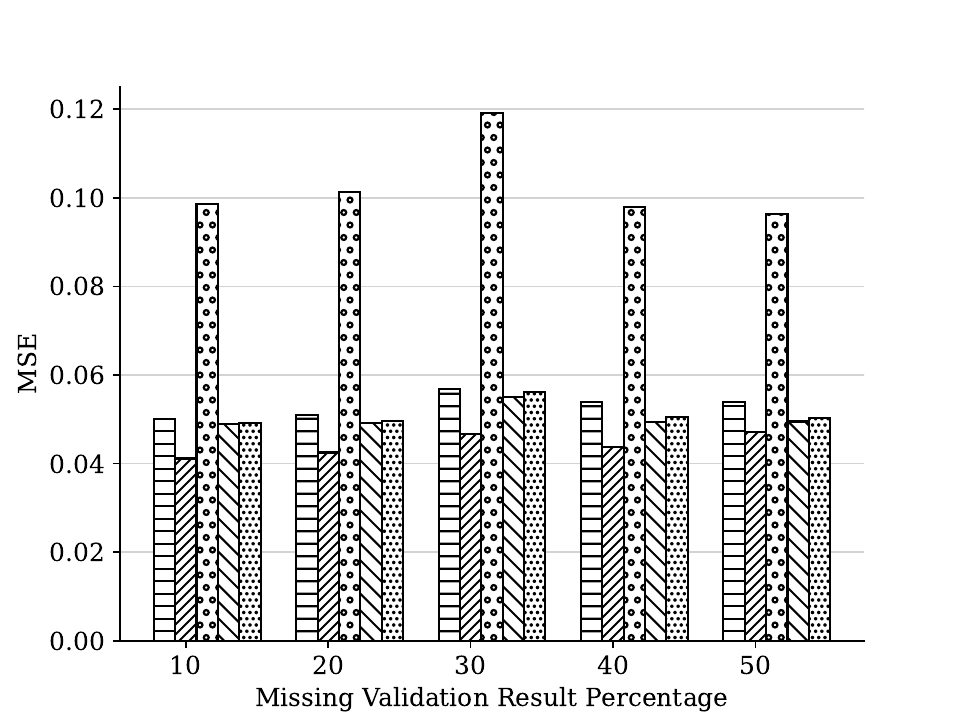}
        \label{fig:ablation_imputation_comparison_one_class_expert}
    }
    \hspace{-22pt}
    \subfigure[]{
        \includegraphics[width=0.26\textwidth]{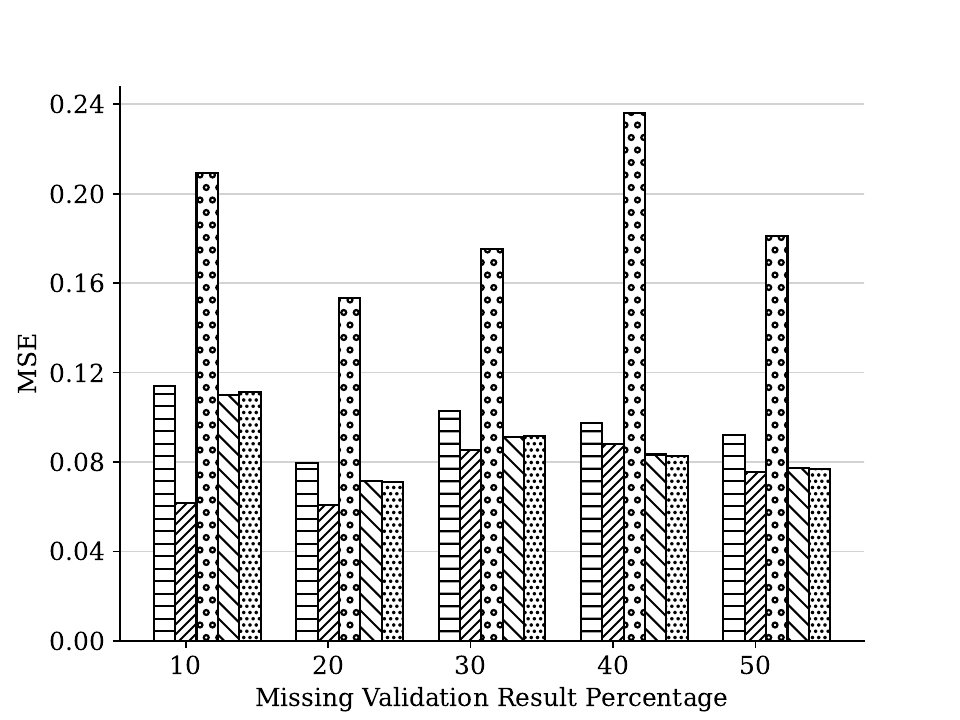}
        \label{fig:ablation_imputation_comparison_sampling_dirichlet}
    }
    \hspace{-22pt}
    \subfigure[]{
        \includegraphics[width=0.23\textwidth]{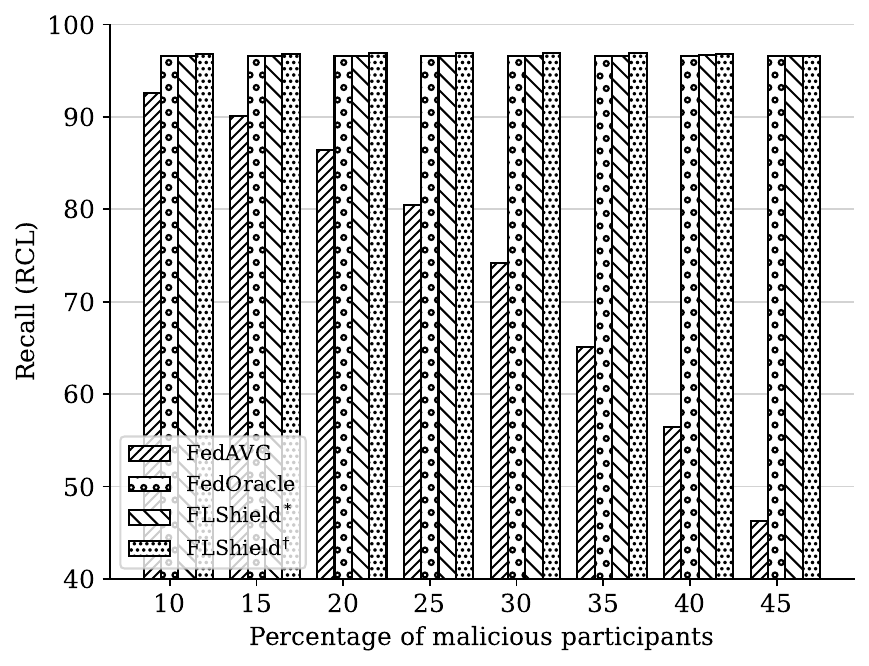}
    \label{fig:mal_pcnt}
    }
    \caption{(a-c) Comparison of imputation techniques' performance with varying missing validation results in different data distribution: (from left to right) IID, one-class-expert, Dirichlet. (d) Effect of varying malicious client participation}
    \label{fig:ablation_imputation}
\end{figure*}

\vspace{-0.1cm}
\subsubsection{Explanation of Design Choices}
\label{ablation:miscellaneous}
\begin{itemize}[leftmargin=0pt, itemindent=15pt]

    \item \emph{Would \textup{\nameofdefense{}} work if the mean of $\mathcal{L}$ is computed instead of minimum before ranking the representative models?} Experimental results presented in Table \ref{tab:ablation_use_mean}\textcolor{green}{(c)} show that using mean instead of minimum leads to an overall loss of RCL, TPR, and TNR. The experiment is run on CIFAR-10 with \nameofdefense{}$^{\dagger}$ as defense and the adversary is launching TLFA.
    
    \item \emph{Does the choice of clustering algorithm matter in \textup{\nameofdefense{}}$^{*}$?} We conduct experiments employing a variety of clustering algorithms, including Agglomerative, K-means, Spectral, and HDBSCAN. To compare their effectiveness, we design a metric named \emph{fidelity score} which is defined in the following way\textemdash{} for each cluster, we first determine whether it has a benign/malicious majority (ground-truth) and then the member count ratio of the majority is computed by dividing the count by the total number of participants in the cluster. Hence, fidelity score is a suitable metric to determine the overall purity of the clusters. The results are presented in Figure \ref{fig:ablation_clustering}. It shows that K-Means have the best overall fidelity score among all  which we use in \nameofdefense{} as the default clustering algorithm.
    
    \item \emph{Does the choice of imputation technique matter in filling out the missing values in the $\mathcal{N}$ matrices?} We undertake an evaluation of different imputation techniques, namely KNNImpute, SimilarityWeightedAveraging, IterativeImpute, Mean, and Median as implemented in \cite{fancyimpute}, by selectively removing a random fraction of the validation results and presenting the findings in Figure \ref{fig:ablation_imputation}. Most algorithms perform well even mean and median where the missing value is simply replaced with mean and median respectively. This again goes to show the congruency of the validation reports. Iterative imputation slightly outperforms the others overall and as such, we use it as the default imputation in \nameofdefense{}.
    
    \item \emph{Why doesn't \textup{\nameofdefense{}} aggregate representative models to obtain the global model?} The representative models selected after outlier removal still include some malicious contribution (see Figure~\ref{fig:ablation_contribution_report4}) and thus may still leave some poisoning footprints in the global model. Experimental results reported in Table~\ref{tab:ablation_aggregate_ensemble}\textcolor{green}{(a)} show that with the presence of malicious contribution of only $8.6\%$, aggregating the representative models instead of the individual local models may increase the backdoor accuracy to $51\%$. 
    
    \item \emph{Does the number of validators selected in each iteration matter?} Table \ref{tab:ablation_num_of_validators}\textcolor{green}{(d)} presents comparison among experiments run with different number of validators ranging from $10$ to $25$ with a step size of $5$. With only $10$ validators, there is a slight drop in performance and in other cases, the TPR is $100\%$ which again demonstrates the robustness of \nameofdefense{}. The experiment is run on the CIFAR-10 dataset against DBA.
\end{itemize}

\subsubsection{Varying Percentage of Malicious Clients}
\label{subsec:impact_of_malicious_clients}
We vary the percentage of malicious clients in the TLFA setting and measure the performance of \nameofdefense{}.
We use F-MNIST dataset for this experiment and vary the percentage of malicious clients from $10\%$ to $45\%$ in steps of $5\%$. Figure \ref{fig:mal_pcnt} shows how the percentage of malicious clients influences \nameofdefense{}$^{*}$ and \nameofdefense{}$^{\dagger}$. 
As shown in the figure, unlike FedAvg, there is no noticeable drop in the performance.

\subsection{Gradient Inversion Attack}
\label{sec:gradient_inversion_attack}
\vspace{-.2cm}
Since sharing gradients in FL may leak private information \cite{gradAttack, gradient_inversion_2020, gradient_inversion_2021, gradient_inversion_iDLG}, we evaluate \nameofdefense{} against a state-of-the-art gradient inversion attack\cite{gradient_inversion_2020} leveraging the attack implemented in \cite{gradAttack}. Note that, in \nameofdefense{}, only the representative gradients are sent to the validators. However, the malicious validators may still try to conduct  gradient inversion attacks on the representative models to extract private local data. To eliminate this concern, we evaluate the robustness of the representative gradients on the CIFAR-10 dataset with ResNet-18 architecture. 
As mentioned in \cite{gradient_inversion_2020}, there are two adversary knowledge assumptions:  (1) private labels and (2) batch normalization statistics.
The detailed attack setting is provided in Appendix \ref{app:attacks}.

Figure \ref{fig:grad_inversion} visualizes the images reconstructed from local gradients (first row), cluster representative models in \nameofdefense{}$^{*}$ (second row), and bijective representative models in \nameofdefense{}$^{\dagger}$. For \nameofdefense{}$^{*}$ we launch the attack under different assumption. With respect to \nameofdefense{}$^{\dagger}$ we launch the attack under the strongest attacker assumption (the attacker has knowledge of private labels and batch normalization statistics) and variate the number of clients. The results show that the reconstructed images with representative gradients in \nameofdefense{} are unrecognizable even in the strongest attack setup. 
\bluesp{
We use the learned perceptual image patch similarity (LPIPS) score \cite{metrics_LPIPS} to measure the performance of the gradient inversion attack in Table \ref{tab:grad_attack_result} in Appendix.
}

\begin{figure}[t]
    \centering
    \subfigure{
        \includegraphics[width=0.47\textwidth]{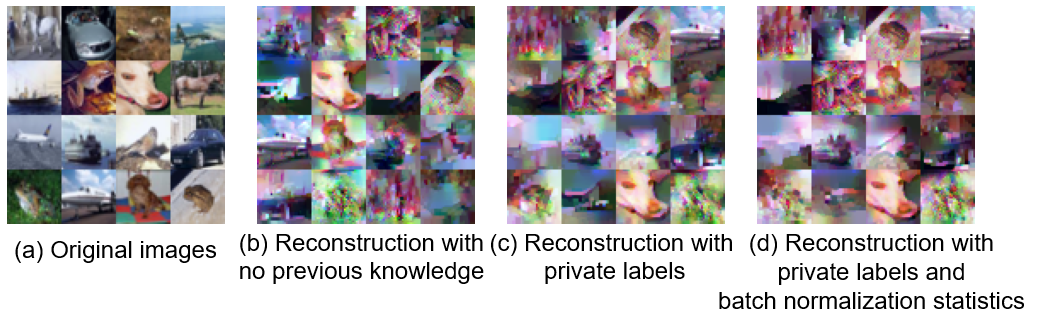}
        \label{fig:gradient_inversion_raw_gradient_16}
    }
    \hspace{-8pt}
    \subfigure{
        \includegraphics[width=0.47\textwidth]{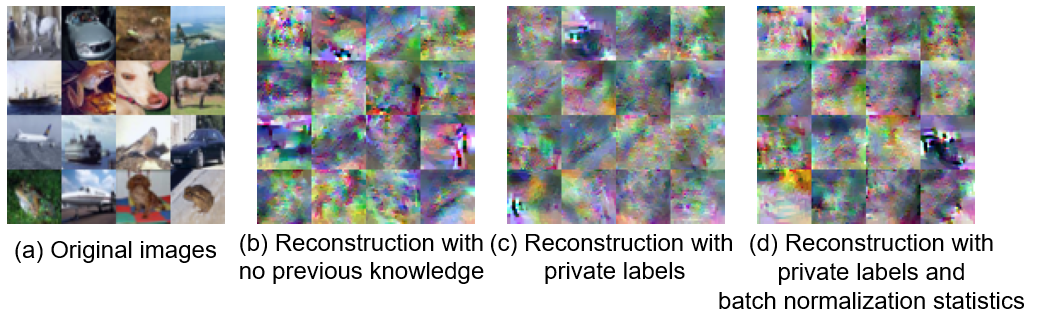}
        \label{fig:gradient_inversion_batch_16}
    }
    \hspace{-8pt} 
    \subfigure{
    \includegraphics[width=0.24\textwidth]{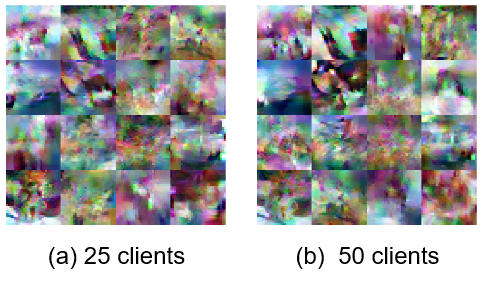}
    \label{fig:gradient_inversion_batch_16_bi}
    }
    \caption{Comparison of images reconstructed from local gradients (first row), representative gradients in \nameofdefense{}$^*$ (second row), and \nameofdefense{}$^{\dagger}$ (third row).}

    \label{fig:grad_inversion}
\end{figure}

\section{Related Work}
\vspace{-0.1cm}
\textbf{Data Poisoning Attack against FL}.
Data poisoning is an adversarial attack that manipulating training datasets by injecting poisoned data to control the behavior of the trained model. This attack have been demonstrated to be successful on many machine learning systems  \cite{svms, featureSelection, neuralNetwork1,trojan, Poisoning_Regression_Learning, poisoning_Influence_Functions, poisoning_Optimal_Training-Set_Attacks, gmm}. 
FL is also vulnerable to data poisoning. In FL, this attack involve maliciously manipulating or poisoning the data contributed by specific clients in order to compromise the overall model's performance \cite{Bag2018, bhagoji, dba, wang2020attack, tolpegin2020data}. Depending on the attacker’s goal, this kind of attacks can be categorized into untargeted attacks \cite{modelpoisoning-gong} and targeted attacks \cite{Bag2018, bhagoji, dba}. Untargeted attacks aim to make the learned model unusable by degrading the performance of the model, while targeted attacks only influence the model's behavior towards misclassifying certain target inputs.

\vspace{0.1cm}
\noindent\textbf{Robust Aggregation Frameworks}.
A number of robust FL aggregation frameworks have been proposed \cite{jebreel2022fldefender, mediantrimmedmean, Blanchard2017, AFA, fltrust, Flame,Bag2018,backdoor_2, shen2016, foolsgold, Rieger_2022, baffle} to mitigate the threat of Byzantine participants. There are mainly two categories of methods: (1) identifying and filtering out malicious clients, and (2) adding noise to local updates.

 Identifying and filtering out malicious clients is a commonly employed method. Some methods \cite{Blanchard2017, gmm, mediantrimmedmean, AFA, shen2016} mitigate malicious clients by removing local updates which are far away from the majority. However, these methods are highly sensitive to outliers in the client updates and may exclude benign clients with significantly different updates. \ignore{In Adaptive Federated Averaging(AFA) \cite{AFA}, cosine similarities between the global update and the local updates are utilized to detect potential malicious model updates. Nonetheless, AFA fails to handle the scenarios where data are non-iid \cite{Flame}.} FLTrust \cite{fltrust}, which makes the assumption that the defender has access to a server-side clean dataset, leverages cosine similarity to identify malicious intent in client updates deviating from a benign server model. However, the challenge lies in assembling such a clean dataset. 

 Certain methods allow clients to act as validators to detect malicious updates. Zhao et al. \cite{zhao2020shielding} propose assessing the integrity of local models using client's local data, marking poorly performing models as malicious, but this method is unable to detect targeted attacks that don't diminish the overall performance of the model. BaFFLe \cite{baffle} lets selected clients to validate the aggregated model but also shares the same problem.

Recent works have introduced many novel statistical techniques to identify abnormal updates. FLDetector \cite{zhang2022fldetector} let the server predict local model updates based on the historical updates and then identify the malicious updates by comparing the predicted updates and the received updates. However, it is not necessary for the malicious clients to keep changing their local data distribution. Deepsight \cite{Rieger_2022} remove the local updates that trained with datasets where a single class dominates the majority of the labels by estimating distribution of clients' local data. Nevertheless, it's easy for an attacker to construct a training set to make the malicious clients evade the detection and the benign clients who only have homogeneous data will suffer from bad model performance. Adding noise to local updates have been explored as a potential option \cite{Flame,Bag2018,backdoor_2} but the noise injection causes utility loss to the trained model.

\ignore{
FLTrust defines a trust score for each client: $TS_{k} = ReLU(\frac{<G_{k},G_{0}>}{||G_{k}||\cdotp||G_{0}||})$, where $G_{0}$ is the server model update, $G_{k}$ is the $i$th client's update. ReLU here is to eliminate the impact of updates in opposite directions. What's more, FLTrust also considers the magnitude of local updates. FLTrust normalizes each local model update such that it has the same magnitude as the server model update: $\bar{G}_{k} = \frac{||G_{0}||}{||G_{k}||}\cdotp G_{k}$, where $\bar{G}_{k}$ is the normalized local model update of the $k$th client. 

}

\vspace{0.1cm}
\noindent\textbf{Gradient Inversion Attacks}.
Gradient Inversion Attack(GIA) \cite{gradient_inversion_2020, gradient_inversion_2021, wei2020framework} aims to reconstruct training samples from gradient for image classification tasks. Adversaries leverage various reconstruction techniques, such as optimization algorithms or machine learning models, to analyze the gradients and infer the underlying training samples. In FLShield representative gradients are sent to clients for validation. Malicious clients may try to reconstruct the training samples from the these gradients.
\vspace{-0.1cm}
\section{Limitations}
\label{sec:limitation}
\vspace{-0.1cm}
One limitation of \nameofdefense{} is that it has been designed to work in FL systems with a classification task.
The extension of \nameofdefense{}'s validation mechanism to other  tasks, e.g., text generation and recommendation systems seem challenging.
It remains a future work to investigate how the insights gained from \nameofdefense{} could be used in those tasks.
In  \nameofdefense{}, all model updates selected post-validation receive equal weight, despite their respective scores. 
We did not integrate an adaptive weighting scheme that might assign a higher weight to models achieving better validation scores. 
Although this has not resulted performance drop in terms of any metric, investigating alternative schemes may be advantageous for an FL system that uses an incentive mechanism to stimulate the provision of high-quality data.
Nevertheless, we leave this exploration as a future work.
\vspace{-0.1cm}
\section{Conclusion}
\vspace{-.1cm}
We introduce a novel validation-based defense framework for Federated Learning (FL) \nameofdefense{} which is adept at resolving the \emph{validation dilemmas} without creating any system vulnerabilities and exhibits robustness against three types of poisoning attacks.
Further, \nameofdefense{} can efficiently defend against two defense-aware attacks, specifically, FA-Adp and FA-Adv. 
We also perform gradient inversion attack experiments, demonstrating that malicious validators are incapable of reconstructing training data from the shared representative models.
This design aligns well with existing privacy precautions against inference attacks. 
We demonstrate that employing a validation process is the sole reliable defense strategy against poisoning attacks in FL. 
Despite our framework being tailored for training classifiers, we anticipate its potential extension to other machine learning tasks, an area we earmark for future research.
\clearpage

\bibliographystyle{acm}
\bibliography{applications}

\begin{thebibliography}{10}

\bibitem{baffle}
{\sc Andreina, S., Marson, G.~A., Mollering, H., and Karame, G.}
\newblock {BaFFLe: Backdoor detection via feedback-based federated learning}.
\newblock In {\em Proceedings - International Conference on Distributed
  Computing Systems\/} (2021), vol.~2021-July.

\bibitem{Bag2018}
{\sc Bagdasaryan, E., Veit, A., Hua, Y., Estrin, D., and Shmatikov, V.}
\newblock {20-AISTATS-C-18-17-How To Backdoor Federated Learning}.
\newblock {\em AISTATS\/} (2018).

\bibitem{bhagoji}
{\sc Bhagoji, A.~N., Chakraborty, S., Mittal, P., and Calo, S.}
\newblock {Analyzing federated learning through an adversarial lens}.
\newblock In {\em 36th International Conference on Machine Learning, ICML
  2019\/} (2019), vol.~2019-June.

\bibitem{svms}
{\sc Biggio, B., Nelson, B., and Laskov, P.}
\newblock Poisoning attacks against support vector machines, 2012.

\bibitem{Blanchard2017}
{\sc Blanchard, P., El~Mhamdi, E.~M., Guerraoui, R., and Stainer, J.}
\newblock {Machine learning with adversaries: Byzantine tolerant gradient
  descent}.
\newblock In {\em Advances in Neural Information Processing Systems\/} (2017),
  vol.~2017-December.

\bibitem{fl_defense_privacy_SMC}
{\sc Bonawitz, K., Eichner, H., Grieskamp, W., Huba, D., Ingerman, A., Ivanov,
  V., Kiddon, C., Kone{\v{c}}n{\`y}, J., Mazzocchi, S., McMahan, B., et~al.}
\newblock Towards federated learning at scale: System design.
\newblock {\em Proceedings of Machine Learning and Systems 1\/} (2019),
  374--388.

\bibitem{fl_aggregation_Practical}
{\sc Bonawitz, K., Ivanov, V., Kreuter, B., Marcedone, A., McMahan, H.~B.,
  Patel, S., Ramage, D., Segal, A., and Seth, K.}
\newblock Practical secure aggregation for privacy-preserving machine learning.
\newblock In {\em proceedings of the 2017 ACM SIGSAC Conference on Computer and
  Communications Security\/} (2017), pp.~1175--1191.

\bibitem{fedavg}
{\sc Brendan~McMahan, H., Moore, E., Ramage, D., Hampson, S., and
  Ag{\"{u}}era~y Arcas, B.}
\newblock {Communication-Efficient Learning of Deep Networks from Decentralized
  Data}.
\newblock {\em Proceedings of the 20th International Conference on Artificial
  Intelligence and Statistics, AISTATS 2017\/} (2 2016).

\bibitem{fl_app_predictive_health}
{\sc Brisimi, T.~S., Chen, R., Mela, T., Olshevsky, A., Paschalidis, I.~C., and
  Shi, W.}
\newblock Federated learning of predictive models from federated electronic
  health records.
\newblock {\em International journal of medical informatics 112\/} (2018),
  59--67.

\bibitem{fltrust}
{\sc Cao, X., Fang, M., Liu, J., and Gong, N.~Z.}
\newblock Fltrust: Byzantine-robust federated learning via trust bootstrapping.
\newblock In {\em 28th Annual Network and Distributed System Security
  Symposium, {NDSS} 2021, virtually, February 21-25, 2021\/} (2021), The
  Internet Society.

\bibitem{neuralNetwork1}
{\sc {Chen}, X., {Liu}, C., {Li}, B., {Lu}, K., and {Song}, D.}
\newblock {Targeted Backdoor Attacks on Deep Learning Systems Using Data
  Poisoning}.
\newblock {\em arXiv e-prints\/} (Dec. 2017), arXiv:1712.05526.

\bibitem{gmm}
{\sc Chen, Y., Su, L., and Xu, J.}
\newblock {Distributed Statistical Machine Learning in Adversarial Settings}.
\newblock {\em ACM SIGMETRICS Performance Evaluation Review 46}, 1 (2019).

\bibitem{dataset_Extending_MNIST}
{\sc Cohen, G., Afshar, S., Tapson, J., and Van~Schaik, A.}
\newblock Emnist: Extending mnist to handwritten letters.
\newblock In {\em 2017 international joint conference on neural networks
  (IJCNN)\/} (2017), IEEE, pp.~2921--2926.

\bibitem{modelpoisoning-gong}
{\sc Fang, M., Cao, X., Jia, J., and Gong, N.~Z.}
\newblock {Local Model Poisoning Attacks to Byzantine-Robust Federated
  Learning}.
\newblock {\em Proceedings of the 29th USENIX Security Symposium\/} (11 2019),
  1623--1640.

\bibitem{foolsgold}
{\sc Fung, C., Yoon, C.~J., and Beschastnikh, I.}
\newblock {The limitations of federated learning in sybil settings}.
\newblock In {\em RAID 2020 Proceedings - 23rd International Symposium on
  Research in Attacks, Intrusions and Defenses\/} (2020).

\bibitem{gradient_inversion_2020}
{\sc Geiping, J., Bauermeister, H., Dr\"{o}ge, H., and Moeller, M.}
\newblock Inverting gradients - how easy is it to break privacy in federated
  learning?
\newblock In {\em Advances in Neural Information Processing Systems\/} (2020),
  H.~Larochelle, M.~Ranzato, R.~Hadsell, M.~Balcan, and H.~Lin, Eds., vol.~33,
  Curran Associates, Inc., pp.~16937--16947.

\bibitem{dataset_LOAN}
{\sc George, N.}
\newblock All lending club loan data, 2019.
\newblock {\em URL https://www.kaggle.com/datasets/zaurbegiev/my-dataset\/}.

\bibitem{fl_app_autonomous_industry4_privacy}
{\sc Hao, M., Li, H., Luo, X., Xu, G., Yang, H., and Liu, S.}
\newblock Efficient and privacy-enhanced federated learning for industrial
  artificial intelligence.
\newblock {\em IEEE Transactions on Industrial Informatics 16}, 10 (2019),
  6532--6542.

\bibitem{fl_defense_privacy_Homomorphic_entity_resolution}
{\sc Hardy, S., Henecka, W., Ivey-Law, H., Nock, R., Patrini, G., Smith, G.,
  and Thorne, B.}
\newblock Private federated learning on vertically partitioned data via entity
  resolution and additively homomorphic encryption.
\newblock {\em arXiv preprint arXiv:1711.10677\/} (2017).

\bibitem{he2016}
{\sc He, K., Zhang, X., Ren, S., and Sun, J.}
\newblock Deep residual learning for image recognition.
\newblock In {\em Proceedings of the IEEE conference on computer vision and
  pattern recognition\/} (2016), pp.~770--778.

\bibitem{fl_defense_privacy_dp}
{\sc Hitaj, B., Ateniese, G., and Perez-Cruz, F.}
\newblock Deep models under the gan: information leakage from collaborative
  deep learning.
\newblock In {\em Proceedings of the 2017 ACM SIGSAC conference on computer and
  communications security\/} (2017), pp.~603--618.

\bibitem{gradAttack}
{\sc Huang, Y., Gupta, S., Song, Z., Li, K., and Arora, S.}
\newblock Evaluating gradient inversion attacks and defenses in federated
  learning.
\newblock In {\em NeurIPS\/} (2021).

\bibitem{Poisoning_Regression_Learning}
{\sc Jagielski, M., Oprea, A., Biggio, B., Liu, C., Nita-Rotaru, C., and Li,
  B.}
\newblock Manipulating machine learning: Poisoning attacks and countermeasures
  for regression learning.
\newblock In {\em 2018 IEEE Symposium on Security and Privacy (SP)\/} (2018),
  pp.~19--35.

\bibitem{jebreel2022fldefender}
{\sc Jebreel, N., and Domingo-Ferrer, J.}
\newblock Fl-defender: Combating targeted attacks in federated learning, 2022.

\bibitem{poisoning_Influence_Functions}
{\sc Koh, P.~W., and Liang, P.}
\newblock Understanding black-box predictions via influence functions, 2017.

\bibitem{konevcny2016federated}
{\sc Kone{\v{c}}n{\`y}, J., McMahan, H.~B., Yu, F.~X., Richt{\'a}rik, P.,
  Suresh, A.~T., and Bacon, D.}
\newblock Federated learning: Strategies for improving communication
  efficiency.
\newblock {\em arXiv preprint arXiv:1610.05492\/} (2016).

\bibitem{dataset_CIFAR}
{\sc Krizhevsky, A., Hinton, G., et~al.}
\newblock Learning multiple layers of features from tiny images.

\bibitem{li2021}
{\sc Li, X., Qu, Z., Zhao, S., Tang, B., Lu, Z., and Liu, Y.}
\newblock Lomar: A local defense against poisoning attack on federated
  learning.
\newblock {\em IEEE Transactions on Dependable and Secure Computing\/} (2021).

\bibitem{trojan}
{\sc Liu, Y., Ma, S., Aafer, Y., Lee, W.-C., Zhai, J., Wang, W., and Zhang, X.}
\newblock Trojaning attack on neural networks.
\newblock In {\em NDSS\/} (2018), The Internet Society.

\bibitem{poisoning_Optimal_Training-Set_Attacks}
{\sc Mei, S., and Zhu, X.}
\newblock Using machine teaching to identify optimal training-set attacks on
  machine learners.
\newblock In {\em Proceedings of the Twenty-Ninth AAAI Conference on Artificial
  Intelligence\/} (2015), AAAI'15, AAAI Press, p.~2871–2877.

\bibitem{minka2000}
{\sc Minka, T.}
\newblock Estimating a dirichlet distribution, 2000.

\bibitem{AFA}
{\sc Muñoz-González, L., Co, K.~T., and Lupu, E.~C.}
\newblock Byzantine-robust federated machine learning through adaptive model
  averaging, 2019.

\bibitem{Flame}
{\sc Nguyen, T.~D., Rieger, P., Chen, H., Yalame, H., Möllering, H.,
  Fereidooni, H., Marchal, S., Miettinen, M., Mirhoseini, A., Zeitouni, S.,
  Koushanfar, F., Sadeghi, A.-R., and Schneider, T.}
\newblock Flame: Taming backdoors in federated learning, 2021.

\bibitem{rfa}
{\sc Pillutla, K., Kakade, S.~M., and Harchaoui, Z.}
\newblock {Robust Aggregation for Federated Learning}.

\bibitem{fl_app_autonomous_vehicles}
{\sc Pokhrel, S.~R., and Choi, J.}
\newblock A decentralized federated learning approach for connected autonomous
  vehicles.
\newblock In {\em 2020 IEEE Wireless Communications and Networking Conference
  Workshops (WCNCW)\/} (2020), IEEE, pp.~1--6.

\bibitem{fl_app_autonomous_vehicles_blockchain}
{\sc Pokhrel, S.~R., and Choi, J.}
\newblock Federated learning with blockchain for autonomous vehicles: Analysis
  and design challenges.
\newblock {\em IEEE Transactions on Communications 68}, 8 (2020), 4734--4746.

\bibitem{fl_app_autonomous_industry4_blockchain}
{\sc Qu, Y., Pokhrel, S.~R., Garg, S., Gao, L., and Xiang, Y.}
\newblock A blockchained federated learning framework for cognitive computing
  in industry 4.0 networks.
\newblock {\em IEEE Transactions on Industrial Informatics 17}, 4 (2020),
  2964--2973.

\bibitem{Rieger_2022}
{\sc Rieger, P., Nguyen, T.~D., Miettinen, M., and Sadeghi, A.-R.}
\newblock {DeepSight}: Mitigating backdoor attacks in federated learning
  through deep model inspection.
\newblock In {\em Proceedings 2022 Network and Distributed System Security
  Symposium\/} (2022), Internet Society.

\bibitem{fl_app_digital_health}
{\sc Rieke, N., Hancox, J., Li, W., Milletari, F., Roth, H.~R., Albarqouni, S.,
  Bakas, S., Galtier, M.~N., Landman, B.~A., Maier-Hein, K., et~al.}
\newblock The future of digital health with federated learning.
\newblock {\em NPJ digital medicine 3}, 1 (2020), 1--7.

\bibitem{rousseeuw1987silhouettes}
{\sc Rousseeuw, P.~J.}
\newblock Silhouettes: a graphical aid to the interpretation and validation of
  cluster analysis.
\newblock {\em Journal of computational and applied mathematics 20\/} (1987),
  53--65.

\bibitem{fancyimpute}
{\sc Rubinsteyn, A., and Feldman, S.}
\newblock fancyimpute: An imputation library for python.

\bibitem{fl_attack_model_poisoning}
{\sc Shejwalkar, V., and Houmansadr, A.}
\newblock Manipulating the byzantine: Optimizing model poisoning attacks and
  defenses for federated learning.
\newblock In {\em NDSS\/} (2021).

\bibitem{shen2016}
{\sc Shen, S., Tople, S., and Saxena, P.}
\newblock Auror: defending against poisoning attacks in collaborative deep
  learning systems.
\newblock {\em Proceedings of the 32nd Annual Conference on Computer Security
  Applications\/} (2016).

\bibitem{backdoor_2}
{\sc Sun, Z., Kairouz, P., Suresh, A.~T., and McMahan, H.~B.}
\newblock Can you really backdoor federated learning?, 2019.

\bibitem{tolpegin2020data}
{\sc Tolpegin, V., Truex, S., Gursoy, M.~E., and Liu, L.}
\newblock Data poisoning attacks against federated learning systems.
\newblock In {\em European Symposium on Research in Computer Security\/}
  (2020), Springer, pp.~480--501.

\bibitem{fl_defense_privacy_hyrbrid_method}
{\sc Truex, S., Baracaldo, N., Anwar, A., Steinke, T., Ludwig, H., Zhang, R.,
  and Zhou, Y.}
\newblock A hybrid approach to privacy-preserving federated learning.
\newblock In {\em Proceedings of the 12th ACM workshop on artificial
  intelligence and security\/} (2019), pp.~1--11.

\bibitem{fl_security_survey_1}
{\sc Truong, N., Sun, K., Wang, S., Guitton, F., and Guo, Y.}
\newblock Privacy preservation in federated learning: An insightful survey from
  the gdpr perspective.
\newblock {\em Computers \& Security 110\/} (2021), 102402.

\bibitem{wang2020attack}
{\sc Wang, H., Sreenivasan, K., Rajput, S., Vishwakarma, H., Agarwal, S., Sohn,
  J.-y., Lee, K., and Papailiopoulos, D.}
\newblock Attack of the tails: Yes, you really can backdoor federated learning.
\newblock {\em Advances in Neural Information Processing Systems 33\/} (2020),
  16070--16084.

\bibitem{fl_defense_privacy_differential_privacy}
{\sc Wei, K., Li, J., Ding, M., Ma, C., Yang, H.~H., Farokhi, F., Jin, S.,
  Quek, T.~Q., and Poor, H.~V.}
\newblock Federated learning with differential privacy: Algorithms and
  performance analysis.
\newblock {\em IEEE Transactions on Information Forensics and Security 15\/}
  (2020), 3454--3469.

\bibitem{wei2020framework}
{\sc Wei, W., Liu, L., Loper, M., Chow, K.-H., Gursoy, M.~E., Truex, S., and
  Wu, Y.}
\newblock A framework for evaluating gradient leakage attacks in federated
  learning, 2020.

\bibitem{featureSelection}
{\sc Xiao, H., Biggio, B., Brown, G., Fumera, G., Eckert, C., and Roli, F.}
\newblock Is feature selection secure against training data poisoning?

\bibitem{dataset_Fashion_mnist}
{\sc Xiao, H., Rasul, K., and Vollgraf, R.}
\newblock Fashion-mnist: a novel image dataset for benchmarking machine
  learning algorithms.
\newblock {\em arXiv preprint arXiv:1708.07747\/} (2017).

\bibitem{dba}
{\sc Xie, C., Huang, K., Pin-Yu, C., and Li, B.}
\newblock {Dba : Distributed Backdoor Attacks}.
\newblock {\em 8th International Conference on Learning Representations,
  {\{}ICLR{\}} 2020\/} (2020).

\bibitem{fl_attack_Inner_Product_Manipulation}
{\sc Xie, C., Koyejo, O., and Gupta, I.}
\newblock Fall of empires: Breaking byzantine-tolerant sgd by inner product
  manipulation.
\newblock In {\em Proceedings of The 35th Uncertainty in Artificial
  Intelligence Conference\/} (22--25 Jul 2020), R.~P. Adams and V.~Gogate,
  Eds., vol.~115 of {\em Proceedings of Machine Learning Research}, PMLR,
  pp.~261--270.

\bibitem{mediantrimmedmean}
{\sc Yin, D., Chen, Y., Ramchandran, K., and Bartlett, P.}
\newblock {Byzantine-robust distributed learning: Towards optimal statistical
  rates}.
\newblock In {\em 35th International Conference on Machine Learning, ICML
  2018\/} (2018), vol.~13.

\bibitem{gradient_inversion_2021}
{\sc Yin, H., Mallya, A., Vahdat, A., Alvarez, J.~M., Kautz, J., and Molchanov,
  P.}
\newblock See through gradients: Image batch recovery via gradinversion, 2021.

\bibitem{fl_defense_privacy_Homomorphic}
{\sc Zhang, C., Li, S., Xia, J., Wang, W., Yan, F., and Liu, Y.}
\newblock $\{$BatchCrypt$\}$: Efficient homomorphic encryption for
  $\{$Cross-Silo$\}$ federated learning.
\newblock In {\em 2020 USENIX annual technical conference (USENIX ATC 20)\/}
  (2020), pp.~493--506.

\bibitem{metrics_LPIPS}
{\sc Zhang, R., Isola, P., Efros, A.~A., Shechtman, E., and Wang, O.}
\newblock The unreasonable effectiveness of deep features as a perceptual
  metric, 2018.

\bibitem{zhang2022fldetector}
{\sc Zhang, Z., Cao, X., Jia, J., and Gong, N.~Z.}
\newblock Fldetector: Defending federated learning against model poisoning
  attacks via detecting malicious clients, 2022.

\bibitem{gradient_inversion_iDLG}
{\sc Zhao, B., Mopuri, K.~R., and Bilen, H.}
\newblock idlg: Improved deep leakage from gradients, 2020.

\bibitem{zhao2020shielding}
{\sc Zhao, L., Hu, S., Wang, Q., Jiang, J., Shen, C., Luo, X., and Hu, P.}
\newblock Shielding collaborative learning: Mitigating poisoning attacks
  through client-side detection, 2020.

\end{thebibliography}

\appendix

\begin{table*}[]
\centering
\scriptsize
\caption{Attack taxonomy}
\resizebox{0.80\textwidth}{!}{
\begin{tabular}{|c|c|c|c|}
\hline
\multirow{5}{*}{Poisoning Attacks}     
& Untargeted Poisoning Attacks               & \multicolumn{2}{c|}{Inner Product Manipulation Attack (IPMA) \cite{fl_attack_Inner_Product_Manipulation} } \\
\cline{2-4}

& \multirow{4}{*}{Targeted Poisoning Attacks} & \multicolumn{2}{c|}{Targeted Label Flipping Attack (TLFA) \cite{tolpegin2020data}}                                                                  \\ \cline{3-4} 
                                                                                 &                                                                                & \multirow{3}{*}{Backdoor Attacks}                                             & Distributed Backdoor Attack (DBA)~\cite{dba} \\ \cline{4-4} 
                                                                                 &                                                                                &                                                                       & Edge-case Backdoor Attack (ECBA) \cite{wang2020attack}   \\ \cline{4-4} 
                                                                                 &                                                                                &                                                                       & Semantic
Backdoor Attack (SBA)~\cite{Bag2018}   \\ \hline

Privacy Inference Attack  &  \multicolumn{3}{c|}{Gradient Inversion Attack~ (GIA)~\cite{gradient_inversion_2020} }                                                            \\ \hline

\multirow{2}{*}{\nameofdefense{}-aware Attacks} & 
\multicolumn{3}{c|}{\nameofdefense{}-aware adaptive attack (FA-Adp)}                                                         \\ \cline{2-4}  & \multicolumn{3}{c|}{\nameofdefense{}-aware advanced attack (FA-Adv)}             
\\ \hline

\end{tabular}
}
\label{tab:attack_taxonomy}
\end{table*}

\begin{table}[t]
    \centering
    \scriptsize
    \caption{FL system parameters}
    \label{tab:fl_system_parameters}
    \begin{tabular}{|c|c|c|c|c|}
        \hline
        \textbf{Parameter} & \textbf{E-MNIST} & \textbf{F-MNIST} & \textbf{CIFAR-10} & \textbf{LOAN} \\
        \hline
        \# of Clients & \multicolumn{3}{|c|}{100} & 51 \\
        \hline
        \# of Clients per Round & \multicolumn{2}{|c|}{25} & 10  & 20 \\
        \hline
        \# of Rounds & \multicolumn{2}{|c|}{150} & 400 & 300 \\
        \hline
        \# of Samples per Client & \multicolumn{2}{|c|}{600} & 500 & Variable \\
        \hline
        \# of Classes & \multicolumn{3}{|c|}{10} & 9 \\
        \hline
        batch size & \multicolumn{4}{|c|}{64} \\
        \hline
        learning rate & \multicolumn{3}{|c|}{0.1} & 0.001\\
        \hline
    \end{tabular}
\end{table}

\begin{table}
    \centering
    \scriptsize
    \caption{Attack Parameters}
    \label{tab:attack_parameters}
    \begin{tabular}{|c|c|c|c|c|}
    \hline
    Parameters* & F-MNIST & E-MNIST & CIFAR-10 & LOAN \\
    \hline
    No. of Malicious Clients & \multicolumn{3}{|c|}{40}  & {20} \\
    \hline
    PSPB & \multicolumn{3}{|c|}{20} & {10} \\
    \hline
    Attack iterations & \multicolumn{3}{|c|}{[36, 150]}  & {[201, 300]}  \\
    \hline
    \multicolumn{5}{r}{* All parameters apply to TLFA, DBA, and IPMA except PSPB for IPMA}
    \end{tabular}
\end{table}

\begin{table}[]
    \centering
    \scriptsize
    \caption{Effectiveness of FLShield in comparison with state-of-the-art defenses against Inner Product Manipulation Attack}
    \label{tab:ipma_result}
    \begin{tabular}{llr}
                \hline
            \multirow{2}{*}{Defense}  &  F-MNIST & CIFAR-10 \\
            \cline{2-3}
                   &   MA  &     MA  \\
            \hline 
            FedOracle  &       85.41 &                    79.08 \\
            FedAvg       &           31.56     &       15.28 \\
              RFA &           87.15 &                 80.41 \\
              AFA        & 81.23                 &       65.77 \\
              FLAME       &  86.59         &      79.62 \\
              FLTrust    &      86.03     &      80.07 \\
              \nameofdefense{}$^*$  &  85.33 &              80.76 \\
              \nameofdefense{}$^{\dagger}$  &       85.13 &                80.76 \\
            \hline
        \end{tabular}
\end{table}

\section{Appendix}
\label{sec:appendix}
\subsection{Details of Experiment Setup}
All experiments have been conducted using the PyTorch framework. The backdoor attack codes and all the existing defenses' codes except FLTrust~\cite{fltrust}, FLAME~\cite{Flame} and AFA~\cite{AFA} were directly used from implementations by Xie et al. \cite{dba}. The TLFA and IPMA attacks, FLTrust, FLAME, and AFA defenses have been re-implemented for comparison purposes.

\subsubsection{Datasets}
\label{app:datasets}
We evaluate the performance of our defense on classification tasks trained on four datasets of whom three are image datasets (F-MNIST \cite{dataset_Fashion_mnist}, E-MNIST \cite{dataset_Extending_MNIST}, CIFAR10 \cite{dataset_CIFAR}) and the rest is a tabular dataset (LOAN \cite{dataset_LOAN}).

\subsubsection{Attacks Considered for \nameofdefense{} Evaluation}
\label{app:attacks}
We first describe the specific attack strategies for each and then provide an overview of the attack parameters used in our experiments.

\vspace{0.1cm}
\noindent\textbf{Inner Product Manipulation Attack~\cite{fl_attack_Inner_Product_Manipulation}}.
This attack assumes the adversary has access to the benign updates of the FL system.
Specifically if the benign updates are $w_{i}^{t}$, for $i \in [c_tn+1, n]$ at iteration $t$ then the malicious clients send the following as the poisoned update:
\begin{equation}
    w_{i}^{t} = -\frac{\epsilon_{att}}{|[c_tn+1, n]|} \sum_{j \in [c_tn+1, n]} w_{j}^{t}
\end{equation}

We set the attack strength $\epsilon_{att}$ to 1.0.
To measure the performance of the attack, we use the MA of the target class as the metric.

\vspace{0.1cm}
\noindent\textbf{Targeted Label Flipping Attack~\cite{tolpegin2020data}}.
\label{subsec:appendix_targeted_label_flipping}
This attack aims to misclassify a specific class of samples (\emph{source} class) to another class (\emph{target} class).
The malicious clients flip the labels of the samples from the source class to the target class in their local dataset and train the malicious local model.
The (source, target) class pairs for the datasets F-MNIST, E-MNIST, CIFAR-10, and LOAN are (coat, shirt), (digit 5, digit 3), (automobile, truck), and (current, charged off) respectively.

\vspace{0.1cm}
\noindent\textbf{Distributed Backdoor Attack \cite{dba}}.
In this attack, the adversary split the trigger pattern into multiple parts, and each client injects one of the partial triggers into a fraction of their training samples.
For our experiments, we use the same trigger pattern as in \cite{dba}, and we split the trigger pattern into 4 parts.
The target class for the datasets F-MNIST, E-MNIST, CIFAR-10, and LOAN are pullover, digit 2, bird, and 'Does not meet the credit policy. Status: Charged Off' respectively.

\vspace{0.1cm}
\noindent\textbf{Edge-case Backdoor Attack \cite{wang2020attack}}.
Edge-case backdoors are generated by altering label data points that, while usually correctly classified by the model, are under-represented, or unlikely to be part of the regular training or test data.
We followed the experimental setup of \cite{wang2020attack}.
The adversary uses images of the planes class from Southwest Airlines as edge-case samples and labels them as truck.

\vspace{0.1cm}
\noindent\textbf{Semantic Backdoor Attack \cite{Bag2018}}.
In this attack, the adversary aims to poison the model with the goal for it to produce an attacker-chosen output on visual semantic features (e.g. green cars).
We followed the experiment setup of \cite{Bag2018} and conduct the image classification task on CIFAR dataset. 
We selected the green car images as backdoor and labeled them as bird.

\vspace{0.1cm}
\noindent\textbf{Gradient Inversion Attack \cite{gradient_inversion_2020}}.
An attacker aims to find the reconstructed samples $x_r$ that minimize the loss function $L_{grad}$ between the gradient from client $k$, $\triangledown G_k(x_k)$, and the reconstructed gradient $\triangledown G(x_r)$, according to the optimization problem $\underset{x_r}{argmin} (L_{grad}(\triangledown G(x_r), \triangledown G_k(x_k))) + R_{aux}(x)$ where $R_{aux}$ represents auxiliary knowledge used for regularization.
We implement the attack proposed in \cite{gradient_inversion_2020} with \cite{gradAttack}. 
The attack necessitates a smaller batch size for success, thus we utilize a batch of 16.
We optimize the attack for 5,000 iterations using Adam, with an initial learning rate of 0.1.

\subsubsection{FL Data Distribution}
For image datasets, We consider an FL system containing 100 clients.
For IID data distribution, we assign each client an equal number of samples from the training set.
The distribution strategies for the two non-IID scenarios are described in section \ref{subsec:non_iid}.
For the LOAN dataset, we split it into 51 segments each corresponding to one of the states in the US. 
The \textit{addr\_state} attribute denotes the state where the loan applicant is from. 
This splitting mechanism provides a \textit{natural} non-IID distribution.

\subsubsection{Models} 
For CIFAR-10 and LOAN dataset, we use a lightweight Resnet-18 model \cite{he2016} and a Soft Decision Tree following the implementation in \cite{dba}.
For F-MNIST and E-MNIST, we train a standard convolutional neural network (CNN) as used in \cite{dba}.

\subsubsection{Experiment Parameters}
The FL system parameters are summarized in Table \ref{tab:fl_system_parameters}.
The attack parameters are given in Table \ref{tab:attack_parameters}.
For \nameofdefense{}$^*$, the minimum and maximum number of clusters $k_1$ and $k_2$ are set to $2$ and $\lfloor{n/2}\rfloor$ respectively.
For validation of \nameofdefense{}, the minimum and maximum number of samples per class $n_1$ and $n_2$ are set to $10$ and $30$ respectively.
For each client, we hold out $30\%$ of its data for the validation task.
However, the resulting reduction of the training data does not impact the performance at all as demonstrated in section \ref{sec:overall_eval}.

\begin{figure}[h]
\scriptsize
    \caption{Clustering-based Representative Generation Algorithm}
    \label{alg:clustering}
    \begin{algorithmic}[1]
        \REQUIRE{
            $\omega_1, \omega_2, \dots, \omega_n \quad \triangleright$ local model updates,\\
            $G_t \quad \triangleright$ the global model
        }
        \ENSURE{
            $m \quad \triangleright$ the mapping of each client to a cluster,\\
            $\mathcal{E}_1, \dots, \mathcal{E}_m \quad \triangleright$ the representative models of the $m$ clusters
        }
        \STATE $\triangleright k_1$ and $k_2$ are the lower and upper bounds of the number of clusters respectively
        \STATE $a_1, \dots, a_m \gets DynamicClustering(\omega_{1:n}, k_1, k_2)  \quad \triangleright  a_i$ is a set containing indices of updates in cluster $i$
        \FOR{each $i$ in $[1, m]$}{
            \STATE $\mathcal{E}_i \gets G_t + \underset{j \in a_i}{Mean} \omega_j$
            \STATE $m_j \gets i \quad \forall j \in a_i$
        }
        \ENDFOR
        \RETURN $m, \mathcal{E}_1, \dots, \mathcal{E}_m$
    \end{algorithmic}
\end{figure}

\begin{figure}[h]
\scriptsize
    \caption{Bijective Representative Generation Algorithm}
    \label{alg:Bijective}
    \begin{algorithmic}[1]
        \REQUIRE{
            $\omega_1, \omega_2, \dots, \omega_n \quad \triangleright$  local model updates,\\
            $G \quad \triangleright$ the global model
        }
        \ENSURE{
            $\mathcal{E}_1, \mathcal{E}_2, \dots, \mathcal{E}_n \quad \triangleright$ the bijective representative models
        }
        \FOR{each $i$ in $[1, n]$}{
            \STATE $\mathcal{E} \gets BijectiveRepresentativeGen(G_t, \omega, \tau)$ 
            \STATE $\triangleright BijectiveRepresentativeGen(.)$ uses equation \ref{eqn:bijective_ensembling} to calculate $\mathcal{E}$
        }
        \ENDFOR
        \RETURN $\mathcal{E}_1, \mathcal{E}_2, \dots, \mathcal{E}_n$
    \end{algorithmic}
\end{figure}

\begin{figure}[h]
\scriptsize
    \caption{Validation Algorithm}
    \label{alg:validation_new}
    \begin{algorithmic}[1]
    \REQUIRE{
        $\{\mathcal{E}_1, \mathcal{E}_2, \dots, \mathcal{E}_m\} \quad \triangleright$ representative models, \\ 
        $G \quad \triangleright$ global model,\\
        $S \quad \triangleright$ set of clients,\\
        $c \quad \triangleright$ number of classes
    }
    \ENSURE{
        $\mathcal{M}_1, \mathcal{M}_2, \dots, \mathcal{M}_m \quad \triangleright \mathcal{M}$ report of each representative model
    }
    \FOR{$i$ in $[1, m]$}{
        \STATE randomly sample $k$ validators $v_1, v_2, \dots, v_k \in S$
        \FOR{$v$ in $v_1, v_2, \dots, v_k$}{
            \STATE send both $G$ and $e$ to current validator
            \STATE calculate $\mathcal{L}(\mathcal{E}, G, v)$ using Algorithm \ref{alg:lipc}
            \STATE send $\mathcal{L}(\mathcal{E}_i, G, v)$ back to the server 
        }
        \ENDFOR
    }
    \ENDFOR
    \STATE calculate $\mathcal{M}(\mathcal{E})$ for each representative model $\mathcal{E}$ using equation (6)
    \STATE calculate $\mathcal{N}(v)$ for each validator $v$ using equation (6)
    \STATE use $Imputation$ to fill the missing values in each $\mathcal{N}$ and $\mathcal{M}$
    \STATE use $OutlierDetection$ on all $\mathcal{N}$ to filter out-of-distribution values
    \STATE use remaining $\mathcal{N}$ to update each $\mathcal{M}$
    \RETURN ($\mathcal{M}_1, \mathcal{M}_2, \dots, \mathcal{M}_m$)
    \end{algorithmic}
\end{figure}

\begin{figure}
\scriptsize
    \caption{LIPC Calculation Algorithm}
    \label{alg:lipc}
    \begin{algorithmic}[1]
        \REQUIRE{
            $e \quad \triangleright$ the representative model to be validated,\\
            $G \quad \triangleright$ the global model at previous iteration,\\
            $v \quad \triangleright$ the validator
        }
        \ENSURE{
            $\mathcal{L} \quad \triangleright$ vector of the representative model
        }
        \STATE $D_1, D_2, \dots, D_c \gets SplitByClass(D) \quad \triangleright D$ is validation dataset of $v$
        \FOR{each $i$ in $[1, c]$}{
            \IF{$|D_i| > n_1$}{
                \STATE $D'_i \gets RandomSample(D_i, min(n_2, D_i))$
                $\quad \triangleright n_1$ and $n_2$ are the lower and upper bounds of the number of samples of each class respectively
                \STATE $\mathcal{L}_i \gets \underset{(x,y) \in D'_i}{Mean} \left(Loss(G(x), y)\right) - \underset{(x,y) \in D'_i}{Mean} \left(Loss(e(x), y)\right)$
            }
            \ELSE{
                \STATE $\mathcal{L}_i \gets nan$
                $\quad \triangleright nan$ (not-a-number) value assigned to the class with less than $n_1$ samples
            }
            \ENDIF
        }
        \ENDFOR
        \RETURN $\mathcal{L}$
    \end{algorithmic}
\end{figure}

\begin{figure}
\scriptsize
    \caption{Filtering Algorithm}
    \label{alg:filtering}
    \begin{algorithmic}[1]
        \REQUIRE{
            $\mathcal{M} \quad \triangleright$ matrices of the representative models
        }
        \ENSURE{
            $\mathbb{I} \quad \triangleright$ indices of the selected representative models
        }
        \STATE $\triangleright$ calculate mean $\mathcal{L}$ of all unfiltered validators for each $\mathcal{M}$
        \STATE $\overline{\mathcal{M}} \gets \underset{v}{Mean}\left(\mathcal{M}[v, :]\right) \quad \forall \mathcal{M}$
        \STATE $\triangleright$ extract minimum value from each $\overline{\mathcal{M}}$ and find the top $50\%$ representative models based on the minimum
        \STATE $\mathbb{I} \gets argsort(min(\overline{\mathcal{M}}_1, \dots, \overline{\mathcal{M}}_m))$
        \RETURN $\mathbb{I}_1, \mathbb{I}_2, \dots, \mathbb{I}_{\lfloor{n/2}\rfloor}$
    \end{algorithmic}
\end{figure}

\subsection{Additional Experiment Results}

\subsubsection{Performance Comparison between the use of different outlier detection algorithm in \nameofdefense{}}
\label{sec:outlier_detector_experiment}
We performed an experiment to compare the performance of three outlier detection algorithms - Local Outlier Factor, Isolation Forest, and Elliptic Envelope - in \nameofdefense{} in all 3 different distribution scenarios. In all scenarios, the TPR is reported to be 100\% and TNR is 86.67\%. This demonstrates that the choice of outlier detection algorithm does not impact the TPR and TNR values.

\subsection{\nameofdefense{}-aware attack Implementation}
\label{appendix:mal_val_attack}

\subsubsection{\nameofdefense{}-Aware Adaptive Attack (FA-Adp)}
Formally, the attacker solves the following optimization problem:
{\fontsize{8}{11}\selectfont
\begin{equation}
    \label{eq:adaptive_obj_1}
    \begin{aligned}
        &\underset{l}{argmax} \left( f_1(l) - \lambda_1 \times f_2(l) - \lambda_2 \times f_3(l) \right) \\
        \mathcal{N}_{mal} &= \{ \mathcal{N}_{j} | \quad j \in  \mathcal{V}_{m} \} \\
        f_1(\mathcal{N}_{mal}) &=  \sum_{ x \in \mathcal{E}_{m}} \left|\underset{cl}{min} \left(\underset{i \in \mathcal{V}_{m}}{Mean} \left(\mathcal{N}_i[x,:]\right) \right) \right|^2 \\
        f_2(\mathcal{N}_{mal}) &= \sum_{ x \in \mathcal{E}_{b}} \left|\underset{cl}{min} \left(\underset{i \in \mathcal{V}_{m}}{Mean} \left(\mathcal{N}_i[x,:]\right) \right) \right|^2 \\
        f_3(\mathcal{N}_{mal}) &= \sum_{i \in \mathcal{V}_{m}}{|\mathcal{N}_i - \underset{j \in \mathcal{V}_{b}}{Mean}(\mathcal{N}_j)|}  \\
    \end{aligned}
\end{equation}
}
In the above equation: $\mathcal{E}_{m}$ is the set of malicious representative models, $\mathcal{E}_{b}$ is the set of benign representative models, $\mathcal{V}_{m}$ is the set of malicious validators, $\mathcal{V}_{b}$ is the set of benign validators,  
$\lambda_1$ and $\lambda_2$ are the regularization parameters for the second and the third term respectively.
$f_1, f_2, f_3$ are the three objectives of the optimization problem.
$f_1$ ($f_2$) refers to the sum of the $\mathcal{L}$ scores of the malicious (benign) representative models. 
Increasing $f_1$ and reducing $f_2$ achieves the infiltration objective.
$f_3$ refers to the sum of the absolute difference between the $\mathcal{L}$ vector of each malicious validator and the mean $\mathcal{L}$ vector of the benign validators.
Reducing $f_3$ achieves the stealth objective.

The optimization problem is solved using the gradient descent algorithm, where the attack updates the $\mathcal{N}_{mal}$ matrices by calculating the partial derivative according to the equation \ref{eq:adaptive_obj_1} repeating this process for 1000 iterations. 
During a grid search over the regularization parameters $\lambda_1$ and $\lambda_2$, the attack simulates the steps of \nameofdefense{} selecting the hyperparameter combination that maximizes the change in the $\mathcal{L}$ score which is then used to craft the malicious $\mathcal{N}$ matrices.

\subsubsection{\nameofdefense{}-aware Advanced Attack (FA-Adv)}
In this strategy, the adversary calculates the $\mathcal{N}$ matrices that meet the infiltration objective with the minimal alteration.
We can safely assume that the $\mathcal{L}$ score given to the malicious representative models is higher than the $\mathcal{L}$ score given to the benign representative models.
The malicious validators in this attack bridge the gap between the $\mathcal{L}$ scores of the malicious and benign representative models by giving a lower score to the benign representative models in the class malicious representative models are underperforming and a higher score to the malicious representative models.
We formalize this strategy in the Algorithm presented in Figure \ref{alg:naive}
\begin{figure}
    \scriptsize
    \caption{FA-Adv Algorithm}
    \label{alg:naive}
    \begin{algorithmic}[1]
        \REQUIRE{
            $\mathcal{N}$ of all the the benign validators \\
            $V_{m}$: malicious validators, 
            $V_{b}$: benign validators,
            $\mathcal{E}_{m}$: malicious validators, 
            $\mathcal{E}_{b}$: benign validators,
        }
        \ENSURE{
            $\mathcal{N}$ for all malicious validators
        }
        \STATE $score_x \gets \underset{cl}{min}\left(\mathcal{E}_x\right) \forall x \in \mathcal{E}_{m} \bigcup \mathcal{E}_{b}$
        \STATE $score_{x'} \gets \underset{x \in \mathcal{E}_{m} \bigcup \mathcal{E}_{b}}{min}(score_x) $
        \STATE $x_1, x_2, \dots, x_m \gets \underset{x \in \mathcal{E}_{m} \bigcup \mathcal{E}_{b}}{argsort}(|score_x - score_{x'}|)$
        \FOR{each $x$ in $\{x_1, x_2, \dots, x_m\}$}{
            \STATE $cl \gets \underset{cl}{argmin}\left(\underset{v \in V_{m} \bigcup V_{b}}{Mean}(\mathcal{E}_x[v,:])\right)$
            \IF{$x$ is benign}{
                \STATE $y \gets \text{benign representative with the highest $\mathcal{L}$ score that will be accepted}$
            }
            \ELSIF{$x$ is malicious}{
                \STATE $y \gets \text{malicious representative with the lowest $\mathcal{L}$ score that will be filtered}$
            }
            \ENDIF
            \STATE $cl' \gets \underset{cl}{argmin}\left(\underset{v \in V_{b}}{Mean}(\mathcal{E}_x[v,:]\right)$
            \STATE $\mathcal{L}(x, v_m)[cl] \gets \frac{\sum_{i \in V_{b} \cup V_{m}}{\mathcal{L}(y, i) [cl']} - \sum_{i \in V_{b}}{\mathcal{L}(x, i) [cl]}}{|V_{b}|} \quad \forall v \in V_m$
        } 
        \ENDFOR
        \RETURN (Updated $\mathcal{N}$ for all malicious validators)
    \end{algorithmic}
\end{figure}

\subsection{Server-side Privacy}
\label{sec:privacy}
\ignore{
Federated learning aims to safeguard privacy by maintaining data and computation on the device itself. Unfortunately, the transmission of gradients in FL remains susceptible to privacy breaches \cite{Nasr_2019, fl_attack_privacy_leakage_from_gradient, fl_attack_privacy_unintended_feature_leakage,fl_privacy_attack_userlevelleakage, fl_privacy_attack_membershipinference, gradAttack} if the central server is malicious. Specifically, insiders at the central server, who can inspect the local updates from a particular client in consecutive training rounds, can steal the training data from gradients. 
}

\textbf{Existing Defenses.}
\ignore{
Several solutions have been proposed to mitigate privacy attacks on the server side, by preventing the central server from inspecting parameters sent from a particular user during the global model training process. 
DP protects privacy by adding noise to the local updates. As a trade-off, adding too much noise will significantly compromise the model's performance. Homomorphic encryption secures the learning process by enabling the central server to perform computation on encrypted data. However, there is a practical limitation due to the tremendous computational overhead. SMC allows the central server to update the global model with aggregated local updates. Nonetheless, SMC-based techniques only work effectively in scenarios of honest participants\cite{fl_security_survey_1}.
}
\bluesp{
Existing defenses employ one of the three following defense strategies:
differential privacy (DP) \cite{fl_defense_privacy_hyrbrid_method, fl_defense_privacy_differential_privacy, fl_defense_privacy_dp}, homomorphic encryption \cite{fl_defense_privacy_Homomorphic_entity_resolution, fl_defense_privacy_Homomorphic} and secure multi-party computation(SMC) \cite{fl_defense_privacy_hyrbrid_method, fl_defense_privacy_SMC, fedavg, fl_aggregation_Practical}. 
DP ensures privacy by adding noise to local updates but can compromise the model's performance if too much noise is added. Homomorphic encryption allows the central server to compute encrypted data, but its practical application is limited by substantial computational overhead. SMC enables global model updates with aggregated local updates but is only effective when participants are honest.\cite{fl_security_survey_1}.
}
\begin{figure}
\includegraphics[width=0.5\textwidth, height=2.5cm]
{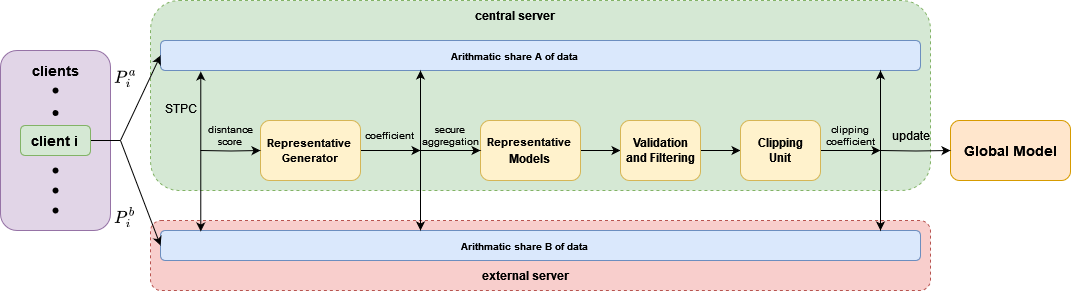}
\caption{Server-side Privacy Strategy}
\label{fig:privateImplementation}
\end{figure}

\textbf{Our Defense.}
Inspired by FLAME \cite{Flame}, we utilize secure two-party computation (STPC) to prevent the threat from malicious central server. The overhead on the runtime caused by STPC is acceptable given it can help maintain privacy.\\ 
The idea is to prevent access to all local updates by a single central server. In order to achieve this goal, a third-party server called External Server is introduced. Figure \ref{fig:privateImplementation} shows the pipeline of our method. In every training round, client $i \in \{1...n\}$ first splits its update $P_i$ into 2 arithmetic shares $P^a_i$ and $P^b_i$, where $P_i = P^a_i + P^b_i$. Then, $P^a_i$ and $P^b_i$ are sent to the central server and the external server separately. This division of data allows both servers to possess partial access to the local updates. Once the central server and the external server receive the partial updates, they can employ STPC to collaboratively generate a representative model and send the generated model to the validators. Last, the servers can clip the local updates based on the output of the validators and update the global model together via STPC.

\begin{figure}[t]
\includegraphics[width=0.5\textwidth, height=2.5cm, keepaspectratio]
{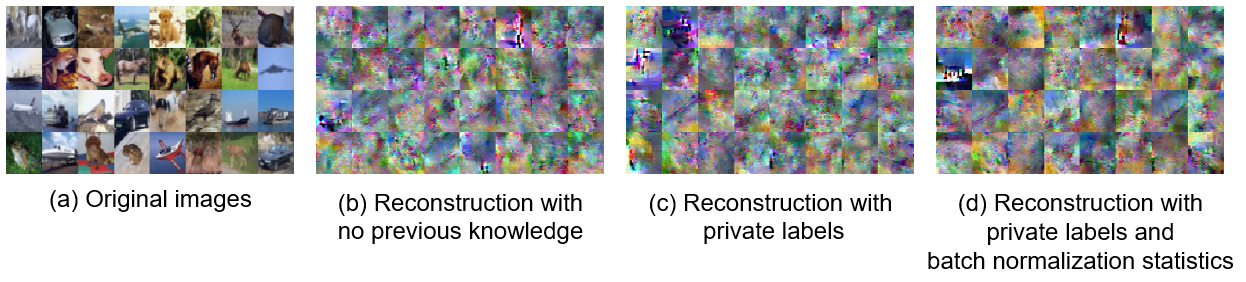}
\caption{Images reconstructed from representative gradients
in \nameofdefense{}$^*$ when batch size is 32}
\label{fig:gradient_inversion_batch_32}
\end{figure}

\begin{table}[h]
\centering
\scriptsize
\begin{tabular}{cccc}
\hline
\multicolumn{1}{c|}{}           & \multicolumn{1}{c|}{None}  & \multicolumn{1}{c|}{Private labels} & \begin{tabular}[c]{@{}c@{}}Private labels\\ + \\ BatchNorm Statistics\end{tabular} \\ \hline
\multicolumn{4}{c}{Attack on a single client's gradient with batch size = 16}                                                                                                                                    \\ \hline
\multicolumn{1}{c|}{Avg. LPIPS $\downarrow$} & \multicolumn{1}{c|}{0.48}  & \multicolumn{1}{c|}{0.48}           & 0.46                                                                               \\
\multicolumn{1}{c|}{Best LPIPS $\downarrow$} & \multicolumn{1}{c|}{0.13}  & \multicolumn{1}{c|}{0.12}           & 0.09                                                                               \\
\multicolumn{1}{c|}{LPIPS std.} & \multicolumn{1}{c|}{0.12}  & \multicolumn{1}{c|}{0.13}           & 0.14                                                                                 \\ \hline
\multicolumn{4}{c}{Attack on cluster representative gradient with batch size = 16}                                                                                                                                              \\ \hline
\multicolumn{1}{c|}{Avg. LPIPS $\downarrow$} & \multicolumn{1}{c|}{0.61}  & \multicolumn{1}{c|}{0.61}           & 0.61                                                                               \\
\multicolumn{1}{c|}{Best LPIPS $\downarrow$} & \multicolumn{1}{c|}{0.53}  & \multicolumn{1}{c|}{0.51}           & 0.48                                                                               \\
\multicolumn{1}{c|}{LPIPS std.} & \multicolumn{1}{c|}{0.03}  & \multicolumn{1}{c|}{0.04}           & 0.04                                                                                 \\ \hline
\multicolumn{4}{c}{Attack on cluster representative gradient with batch size = 32}                                                                                                                                              \\ \hline
\multicolumn{1}{c|}{Avg. LPIPS $\downarrow$} & \multicolumn{1}{c|}{0.60}  & \multicolumn{1}{c|}{0.59}           & 0.60                                                                               \\
\multicolumn{1}{c|}{Best LPIPS $\downarrow$} & \multicolumn{1}{c|}{0.46}  & \multicolumn{1}{c|}{0.50}           & 0.45                                                                               \\
\multicolumn{1}{c|}{LPIPS std.} & \multicolumn{1}{c|}{0.03}  & \multicolumn{1}{c|}{0.03}           & 0.04                                                                               \\ \hline                                                                                                                    
\end{tabular}
\caption{We evaluate the gradient inversion attack against cluster representative gradient in \nameofdefense{}$^*$ on a subset of 50 CIFAR-10 images. ($\downarrow$: lower values suggest more privacy leakage). The averaged and the best results for the metric of reconstruction quality are provided as average-case and worst-case privacy leakage.}
\label{tab:grad_attack_result}
\end{table}

\IEEEpeerreviewmaketitle

\end{document}